\documentclass[a4paper,10pt]{article}
\pdfoutput=1 
\usepackage{jcappub} 

\usepackage{graphics}
\usepackage{graphicx}
\usepackage[T1]{fontenc} 
\usepackage{footnote}
\usepackage{hyperref}
\usepackage{bm}
\usepackage{amsmath}
\def\be{\begin{equation}}
\def\ee{\end{equation}}
\def\ba{\begin{eqnarray}}
\def\ea{\end{eqnarray}}
\def\dd{\textrm{d}}

\title{Probing primordial features with next-generation photometric and radio surveys}

\author[a,b,c,d]{M.~Ballardini,}
\author[c,d]{F.~Finelli,}
\author[a,e]{R.~Maartens,}
\author[b,f,c]{L.~Moscardini}

\affiliation[a]{Department of Physics \& Astronomy, University of the Western Cape, Cape Town 7535, South Africa}
\affiliation[b]{Dipartimento di Fisica e Astronomia, Alma Mater Studiorum 
Universit\`a di Bologna, Via Gobetti 93/2, I-40129 Bologna, Italy}
\affiliation[c]{INAF/OAS Bologna, via Gobetti 101, I-40129 Bologna, Italy}
\affiliation[d]{INFN, Sezione di Bologna, Via Berti Pichat 6/2, I-40127 Bologna, Italy}
\affiliation[e]{Institute of Cosmology \& Gravitation, University of Portsmouth, Portsmouth PO1 3FX, UK}
\affiliation[f]{INAF/OAS Bologna, via Gobetti 93/3, I-40129 Bologna, Italy}

\emailAdd{mario.ballardini@gmail.com}
\emailAdd{finelli@iasfbo.inaf.it}
\emailAdd{roy.maartens@gmail.com}
\emailAdd{lauro.moscardini@unibo.it}

\abstract{
We investigate the possibility of using future photometric and radio surveys to constrain the power 
spectrum of primordial fluctuations that is predicted by inflationary models with a violation of the 
slow-roll phase. We forecast constraints with a Fisher analysis on the amplitude of the parametrized 
features on ultra-large scales, in order to assess whether these could be distinguishable over the 
cosmic variance.
We find that the next generation of photometric and radio surveys has the potential to test these 
models at a sensitivity better than current CMB experiments and that the synergy between galaxy and 
CMB observations is able to constrain models with many extra parameters.
In particular, an SKA continuum survey with a huge sky coverage and a flux threshold of a few $\mu$Jy 
could confirm the presence of a new phase in the early Universe at more than 3$\sigma$.
}

\begin{document}
\maketitle
\flushbottom

\section{Introduction}
\label{sec:intro}

Next-generation spectroscopic, photometric and radio galaxy surveys will allow us to map the Universe 
on the very largest scales -- and thus probe the physics of the primordial fluctuations, as well as 
ultra-large scale general relativistic effects on galaxy observations. 
Several papers have quantified how well we will be able to constrain primordial non-Gaussianity and 
relativistic effects in the galaxy power spectrum, using LSST, SKA and other future surveys (see, e.g. 
\cite{Yoo:2012se,Camera:2013kpa,Camera:2014bwa,Camera:2014sba,Camera:2015yqa, Raccanelli:2015vla, Alonso:2015uua, Baker:2015bva, Alonso:2015sfa, Fonseca:2015laa, Raccanelli:2015oma, Bull:2015lja, DiDio:2016ykq, Alonso:2016suf, Lorenz:2017iez}).

Another interesting target for future surveys is the possibility to refine our knowledge of the 
primordial power spectrum and to carefully investigate the statistical significance of the deviations 
from a simple power law for density fluctuations, that are compatible with the {\em Planck} and WMAP 
CMB temperature power spectrum. 
These deviations from a simple power law  can be easily accommodated in models of inflation with 
temporary violation of the slow-roll conditions.
At present, no inflationary model that fits these features has been found to be preferred at a 
statistically significant level from CMB data 
(see, e.g. \cite{Peiris:2003ff,Covi:2006ci,Planck:2013jfk,Benetti:2013cja,Miranda:2013wxa,Easther:2013kla,Chen:2014joa,Achucarro:2014msa,Hazra:2014goa,Hazra:2014jwa,Hu:2014hra,Ade:2015lrj,Gruppuso:2015zia,Gruppuso:2015xqa,Hazra:2016fkm,Torrado:2016sls}).

The situation improves if suitable data in addition to the CMB temperature are available. Better 
CMB E-mode polarization measurements have been highlighted as a possible way to constrain  primordial 
features with high confidence \cite{Finelli:2016cyd,Hazra:2017joc}.
Galaxy surveys provide a unique opportunity to improve our current understanding about these possible 
anomalies; see for instance 
\cite{Huang:2012mr,Chen:2016vvw,Chen:2016zuu,Ballardini:2016hpi,Xu:2016kwz, Munoz:2016owz, Pourtsidou:2016ctq,Fard:2017oex,Palma:2017wxu,LHuillier:2017lgm}. 

Here we focus on constraining primordial features using future photometric and radio galaxy surveys 
that cover a huge volume of the Universe, in order to access the ultra-large scales where primordial 
features leave an imprint. As examples of such surveys, we use two experiments that are being 
constructed:
\begin{itemize}\item
The Large Synoptic Survey Telescope\footnote{\href{http://www.lsst.org/}{http://www.lsst.org/}} 
(LSST) is the widest ($\sim18,000$ deg$^2$) and deepest ($r_{\rm AB}\sim27.5$) photometric survey 
planned in the foreseeable future, with a sample of $\sim10$ billion galaxies. (See \cite{Zhan:2017uwu}.)
\item
The Square Kilometre Array\footnote{\href{http://www.skatelescope.org/}{http://www.skatelescope.org/}} 
(SKA) plans to conduct the widest ever spectroscopic surveys, using the 21cm emission line of HI: 
with intensity mapping in SKA1-MID ($\sim25,000$ deg$^2$ out to $z\sim 3$), and with a galaxy survey 
in SKA2-MID ($\sim30,000$ deg$^2$, $\sim1$ billion galaxies out to $z\sim2$). In addition, using the 
radio continuum emission, it will detect  a huge number of galaxies out to $z\sim 5$, but without 
redshift information. (See \cite{Maartens:2015mra,Abdalla:2015kra,Santos:2015bsa,Jarvis:2015asa}.)
\end{itemize}

The paper is organized as follows. In section~\ref{sec:two} we describe the observables, the 
forecasting methodology and the survey specifications that we used. In section~\ref{sec:three} we 
describe the parametrized features models used in our forecasts. 
We present our results in section~\ref{sec:four} and draw conclusions in section~\ref{sec:conclusion}.

\section{Large-Scale Structure Power Spectra}
\label{sec:two}

\subsection{Galaxy power spectrum}
Galaxies trace the invisible cold dark matter (CDM) distribution and then we can estimate the matter 
power spectrum and extract information on the underlying power spectrum of primordial fluctuations.
We measure galaxy positions in angular and redshift coordinates and not the position in comoving 
coordinates, i.e. the true galaxy power spectrum is not a direct observable.
We use a model for the observed galaxy power spectrum based on 
\cite{Seo:2003pu,Song:2008qt,Wang:2012bx}:
\be
\label{eqn:Pgal}
P_{\rm g}(k^{\rm ref}_\perp, k^{\rm ref}_\parallel,z) = 
\left[ \frac{D^{\rm ref}_{\rm A}(z)}{D_{\rm A}(z)} \right]^2
\frac{H(z)}{H^{\rm ref}(z)} b_{\rm g}^2(z)P_{\rm dw}(k,\mu,z)\,
G_{\rm FoG}(k,\mu,z)\exp\big[-k^2\mu^2\sigma_{r,z}^2\big]
+{{\cal N}_{\rm gal}}(z) \,,
\ee
where $H(z) = \dot{a}/a$ is the Hubble parameter, $D_\mathrm{A} = r(z)/(1+z)$ is the angular diameter 
distance, $r(z)$ is the comoving distance, $b_\mathrm{g}(z)$ is the large-scale galaxy bias, 
$k^2=k_\perp^2+k_\parallel^2$ and $\mu=k_\parallel/k=\hat{\bm{r}}\cdot\hat{\bm{k}}$. This is connected 
to the true galaxy power spectrum via a coordinate transformation \cite{Alcock:1979mp}:
\be
k^{\rm ref}_\perp = \frac{D_{\rm A}(z)}{D^{\rm ref}_{\rm A}(z)}\,k_\perp\,,
\qquad k^{\rm ref}_\parallel=\frac{H^{\rm ref}(z)}{H(z)}\,k_\parallel\,.
\ee
In \eqref{eqn:Pgal}, ${\cal N}_{\rm gal}$ is the shot noise and we model the redshift-space distortions 
(RSD) as:
\be
G_{\rm FoG}(k,\mu,z) = \frac{(1+\beta\mu^2)^2}{1+k^2\mu^2\sigma^2_{r,p}/2}\,,\quad \beta={f(k,z)\over b_{\rm g}(z)}\,, \quad  f(k,z)= {\dd\ln {D}(k,z)\over \dd\ln a}
\ee
where $f$ is the growth rate.
Here the numerator is the linear RSD \cite{Kaiser:1987qv,Hamilton:1997zq}, which takes into account 
the enhancement due to large-scale peculiar velocities. The Lorentzian denominator models the nonlinear 
damping due to small-scale peculiar velocities, where $\sigma_{r,p}$ is the distance dispersion:
\be
{\sigma_{r,p}(z)} = \frac{\sigma_{p}(z)}{H(z)a(z)} \,, 
\ee
corresponding to the physical velocity dispersion $\sigma_p$. We choose a value of 
$\sigma_p = 290$\,km/s as our fiducial \cite{Wang:2012bx}.
An additional exponential damping factor is added to account for {the error $\sigma_z$} in the 
determination of the redshift of sources, where:
\be
\sigma_{r,z}(z) = \frac{\partial r}{\partial z} \sigma_z = \frac{c}{H(z)}\sigma_z \,.
\ee
Finally, the smearing of the BAO feature is modeled by using the dewiggled matter power spectrum:
\be
P_\mathrm{dw}(k,\mu,z) = P_\mathrm{nw}(k,\mu,z) + \Big[P_\mathrm{m}(k,\mu,z) - P_\mathrm{nw}(k,\mu,z)\Big]
\exp\Big[-{g_\mu k^2 \over 2k_*^2}\Big] ,
\ee
where  $P_{\rm nw}$ is the no-wiggle power. The damping along the line-of-sight is described by:
\be
g_\mu(k,\mu,z) = D^2({k},z) \left\{1-\mu^2+\mu^2\big[1+f({k},z)\big]^2\right\} \,,
\ee
and we take $k_* \simeq 0.12\,h/$Mpc, corresponding to the conservative case with no reconstruction 
\cite{Wang:2012bx}.

\subsection{Intensity mapping power spectrum}
Detecting individual galaxies in an HI galaxy redshift survey requires very high sensitivity. 
In Phase 1 of the SKA, the survey  will cover only $\sim 5,000\,$deg$^2$ out to 
$z\sim 0.6$~\cite{Abdalla:2015kra}. For this reason, we only forecast for the SKA HI galaxy 
redshift survey in Phase 2, which will cover $\sim 30,000\,$deg$^2$ out to $z\sim 2$. 
However, there is a way in Phase 1 to achieve very high sky and redshift coverage, but at the 
cost of not detecting individual galaxies.
This is the intensity mapping method: the total HI emission in each pixel is used to give a brightness 
temperature map of the large-scale fluctuations in HI galaxy clustering (with very accurate redshifts)
~\cite{Battye:2004re,Wyithe:2007gz,Chang:2007xk,Santos:2015bsa}.

The flux density measured is converted into an effective brightness temperature of the HI emission, 
which can be split into a homogeneous and a fluctuating part \cite{Bull:2014rha}:
\be \label{imt}
T_{\rm b}=\bar{T}_{\rm b}\big(1+\delta_{\rm HI}\big)\,,\qquad\bar{T}_{\rm b} \approx 566h\frac{\Omega_{\rm HI}(z)}{0.003}(1+z)^2\frac{H_0}{H(z)}
\ \mu\text{K} \,.
\ee
Here $(1+z)^3 \Omega_{\rm HI}(z) = 8\pi G\rho_{\rm HI}(z)/(3H_0^2)$ is the comoving HI density parameter.
We expect HI to be a biased tracer of the CDM distribution, just as galaxies are, because the neutral 
hydrogen content of the Universe is expected to be localized within the galaxies after reionization. 
In real space, $\delta_{\rm HI}=b_{\rm HI}\delta_{\rm m}$, so that \eqref{eqn:Pgal} is modified as 
follows:
\be
\label{eqn:Phi}
P_{\rm HI}(k^{\rm ref}_\perp, k^{\rm ref}_\parallel,z) = 
\left[ \frac{D^{\rm ref}_{\rm A}(z)}{D_{\rm A}(z)} \right]^2
\frac{H(z)}{H^{\rm ref}(z)}\,\bar{T}_b^2(z) b_{\rm HI}^2(z)P_{\rm dw}(k,\mu,z)\,
G_{\rm FoG}(k,\mu,z)\exp\big[-k^2\mu^2\sigma_{r,z}^2\big]
+{\cal N}_{\rm HI}(z) \,,
\ee
where ${\cal N}_{\rm HI}$ is the intensity mapping noise (see below).

\subsection{Fisher forecast formalism}
We follow the same approach as~\cite{Ballardini:2016hpi} (see also~\cite{Tegmark:1997rp,Seo:2003pu}). 
The Fisher matrix for the observed matter power spectrum, for a redshift $z_{\rm i}$ at the centre of 
the i-th bin, is given by:
\be
\label{eqn:fisher}
{\bf{\cal F}}_{\alpha\beta}^{\rm XX}(z_{\rm i}) = \sum_{k,\mu} 
\frac{\partial \ln P_{\rm X}(k,\mu,z_{\rm i})}{\partial \theta_\alpha}\bigg|_{\bar{\theta}}
\,\big[\text{Cov}_{\bf k}(z_{\rm i})\big]^{-1}\,
\frac{\partial \ln P_{\rm X}(k,\mu,z_{\rm i})}{\partial \theta_\beta}\bigg|_{\bar{\theta}} \,,
\ee
where X\,=\,g or HI and
\be
\text{Cov}_{\bf k}(z_{\rm i}) = 
\frac{(2\pi)^2}{k^2 \Delta k\Delta \mu}\frac{1}{V_{\rm eff}(k,\mu,z_{\rm i})} \,.
\ee
We consider 10 bins in $\mu$ between 0 and 1 with $\Delta \mu = 0.1$. 
The effective bin volume is given in terms of the {comoving} bin volume $V_{\rm surv}(z_{\rm i})$ by 
\cite{Feldman:1993ky}:
\ba
\label{eqn:Veff}
V_{\rm eff}(k,\mu,z_{\rm i}) &\simeq& V_\text{surv}(z_{\rm i}) 
\left[\frac{P_{\rm X}(k,\mu,z_{\rm i})}{P_{\rm X}(k,\mu,z_{\rm i})
+{\cal N}_{\rm X}(k,\mu,z_{\rm i})}\right]^2 \,,\\
V_{\rm surv}(z_{\rm i}) &=& \frac{4 \pi f_{\rm sky}}{3}
\left[r^3\Big(z_{\rm i}+\frac{\Delta z}{2}\Big)
- r^3\Big(z_{\rm i}-\frac{\Delta z}{2}\Big)\right] \,.
\ea
The weighting factor in $V_{\rm eff}$ accounts for the varying sensitivity of an experiment to 
different Fourier modes. $P_{\rm X}$ are given in~\eqref{eqn:Pgal} and \eqref{eqn:Phi}, 
and ${\cal N}_{\rm X}$ are given in~\eqref{eqn:Ngal} and \eqref{eqn:Nhi} (see below in 
section~\ref{sec:specifications}).

The full set of parameters $\theta_\alpha$ includes:\\ the standard cosmological parameters $\omega_c$, $\omega_b$, $h$, $n_s$; 
\be \label{param}
\big\{H, D_\mathrm{A}, \log\left(f\sigma_8\right)\big\}_{z_{\rm i}},  
\quad \big\{\log\left(b_{\rm g}\sigma_8\right)~{\rm or}~\log\left(\bar{T}_{\rm b}b_{\rm HI}\sigma_8\right),\sigma_z/(1+z), {\cal N}_\mathrm{X}\big\}_{z_{\rm i}},~ k_* \,;
\ee
and the parameters of the primordial feature models (see below, 
section~\ref{sec:three}).
In the first set of \eqref{param}, we have the Hubble parameter, angular diameter distance and 
linear growth (describing anisotropies in the power spectrum), in each redshift bin. The second set 
contains the nuisance parameters arising from the models for the bias of galaxies or intensity mapping, the photo-$z$ error, and the noise residual. In each redshift bin, the fiducial values of the nuisance parameters are determined by the models described in the text, and our analysis takes account of errors on these fiducial values. The final parameter $k_*$ in \eqref{param} models the nonlinear RSD effect.
After marginalizing over the nuisance parameters, we project the redshift-dependent parameters on 
the final set of cosmological parameters $\omega_c$, $\omega_b$, $h$, $n_s$, 
$\log\left(10^{10}A_s\right)$ and the additional primordial-feature parameters.

The Fisher matrix for CMB angular power spectra (temperature and E-mode polarization) 
is~\cite{Knox:1995dq,Jungman:1995bz,Seljak:1996ti,Zaldarriaga:1996xe,Kamionkowski:1996ks}:
\be
{\cal F}_{\alpha\beta}^\mathrm{CMB} = \sum_{\ell} \sum_{X,Y} 
\frac{\partial C_\ell^{X}}{\partial\theta_\alpha}\bigg|_{\bar{\theta}}\,
\big[\mathrm{Cov}_\ell\big]^{-1}_{XY}\, 
\frac{\partial C_\ell^{Y}}{\partial\theta_\beta}\bigg|_{\bar{\theta}}\,,
\ee
where $X,\,Y = TT,\,TE,\,EE$ and the covariance matrix is:
\be
\mathrm{Cov}_\ell = \frac{2}{(2 \ell+1) f_\mathrm{sky}}
\begin{bmatrix}
\big(\bar{C}_\ell^{TT}\big)^2 & \big(\bar{C}_\ell^{TE}\big)^2 & \bar{C}_\ell^{TT}\bar{C}_\ell^{TE} \\ && \\
\big(\bar{C}_\ell^{TE}\big)^2 & \big(\bar{C}_\ell^{EE}\big)^2 & \bar{C}_\ell^{EE}\bar{C}_\ell^{TE} \\ & & \\
~\bar{C}_\ell^{TT}\bar{C}_\ell^{TE}~ & ~\bar{C}_\ell^{EE}\bar{C}_\ell^{TE}~ & ~\Big\{\bar{C}_\ell^{TT}\bar{C}_\ell^{EE}+\big(\bar{C}_\ell^{TE}\big)^2\Big\}/2~
\end{bmatrix}\,.
\ee
Here $\bar{C}_\ell^{X}$ is the sum of the theoretical $C_\ell^{X}$ and the effective noise 
$N^{X}_{\ell}$, which is the inverse noise weighted combination of the instrumental noise 
convolved with the beams of different frequency channels \cite{Ballardini:2016hpi}.
For the CMB the full set of parameters $\theta_\alpha$ includes $\omega_c$, $\omega_b$, $h$, $n_s$, 
$\log\left(10^{10}A_s\right)$, $\tau$ and the extra primordial-feature parameters. 
We marginalize over the optical depth $\tau$ before combining the CMB Fisher matrix with the Fisher 
matrix of the galaxy/ intenisy mapping power spectrum.
We adopt the specifications denoted as CMB-1 in \cite{Ballardini:2016hpi}, which reproduce 
uncertainties for standard cosmological parameters similar to those which can be obtained by 
{\em Planck}.

\subsection{Survey specifications}
\label{sec:specifications}

\begin{figure}[h!]
\centering
\includegraphics[width=\textwidth]{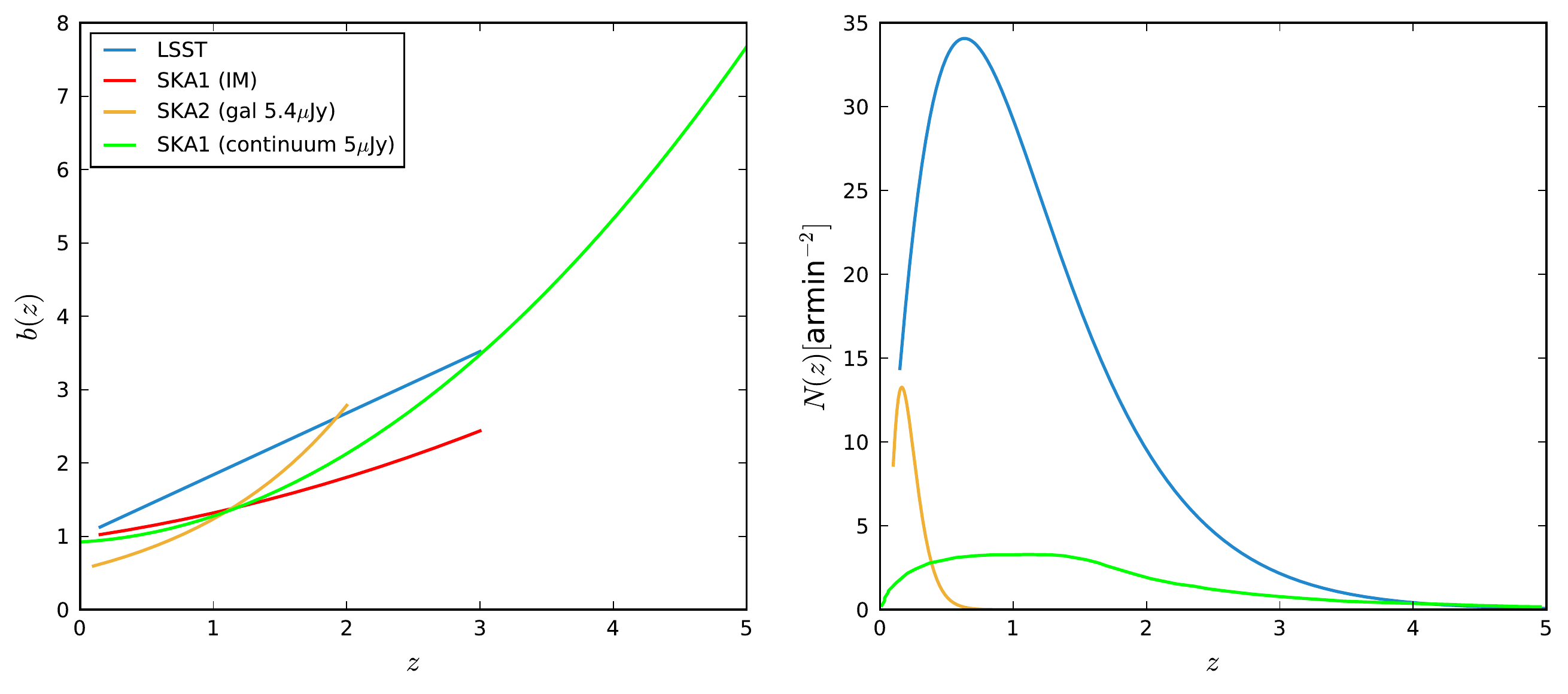}
\caption{Linear clustering bias ({\em left}) and number density of galaxies ({\em right}) for the  
surveys considered. (We do not show the temperature distribution \eqref{imt} of the SKA1 IM.) 
}
\label{fig:spec}
\end{figure}

For the relevant survey specifications and properties of the target galaxies in LSST and SKA1, we have used the most up to date publications that we are aware of.\\

\noindent{\em LSST photometric survey:}\\
We assume a  single-tracer survey over 18,000\,deg$^2$ ($f_{\rm sky} \simeq 0.44$) in the redshift 
range $0.15\le z\le 3.0$. The linear clustering bias and redshift distribution are~\cite{Abell:2009aa}:
\be
b_{\rm g}(z) = 1+0.84z\,,\qquad {\dd N\over\dd z} \propto z^\alpha 
\exp\Big[-\Big({z\over z_0}\Big)^\beta\Big] \,,
\ee 
with $\alpha=2$, $\beta=1$ and $z_0=0.5$ (normalized to have 50 galaxies arcmin$^{-2}$).
The true galaxy distribution is multiplied by a Gaussian photometric redshift error 
distribution~\cite{Ma:2005rc}:
\ba
n_{\rm i}(z) &=& \int^{z_{\rm i}+\Delta z_i}_{z_{\rm i}} \dd z'\, n(z)p(z'|z) \,,\\
p(z_{\rm ph}|z) &=& \frac{1}{\sqrt{2\pi\sigma_{ z}^2}}
\exp\left[-\frac{(z-z_{\rm ph}+z_\mathrm{bias})^2}{2\sigma_{z}^2}\right] \,.
\ea
The redshift uncertainty is $\sigma_{ z} = \sigma_{ z0}(1+z)$, with a conservative value of 
$\sigma_{ z0}=0.05$. We take $z_{\rm bias} = 0$, since any photometric redshift bias known a priori 
can be removed~\cite{Abell:2009aa}.

The noise variance per steradian in the i-th redshift bin is:
\be
\label{eqn:Ngal}
{\cal N}_{\rm gal} = \frac{1}{n_{\rm i}(z)} \,.
\ee
\\

\noindent{\em SKA1 HI intensity mapping (IM):}\\
HI IM surveys will be performed using both interferometer and single-dish modes.
The former has very good angular resolution but is limited to small scales (except at high redshift), 
so the single-dish mode is the most efficient way to probe cosmological scales~\cite{Santos:2015bsa}. 
For SKA1-MID, we assume $t_{\rm tot} = 10^4\,$hours observing over 25,000\,deg$^2$ 
($f_{\rm sky} \simeq 0.60$) in a redshift range $0.15\le z\le 3$ ($1050\geq\nu\geq350\,$MHz). 
We use the fitting formulas~\cite{Santos:2017qgq}:
\ba
b_{\rm HI}(z) &=& \frac{b_{{\rm HI}}(0)}{0.677105} \Big[6.6655 \times 10^{-1} + 1.7765 \times 10^{-1}\, z + 5.0223 \times 10^{-2}\,  z^2\Big] \,, \\
\Omega_{\rm HI}(z) &=& \frac{\Omega_{{\rm HI}}(0)}{0.000486} \Big[4.8304\times 10^{-4} + 3.8856 \times 10^{-4}\, z - 6.5119 \times 10^{-5}\, z^2\Big] \,, \\
\bar{T}_b(z) &=& 5.5919 \times 10^{-2} + 2.3242 \times 10^{-1}\, z - 2.4136 \times 10^{-2}\, z^2\,\,{\rm mK} \,,
\ea
where: 
$\Omega_{\rm HI}(0) = 4.86\times 10^{-4}$ and $\Omega_{\rm HI}(0)b_{\rm HI}(0) = 4.3\times 10^{-4}$.

Assuming scale-independence and no correlation between the noise in different frequency channels, 
the noise variance per steradian in the i-th frequency channel is~\cite{Bull:2015lja}:
\ba
\label{eqn:Nhi}
{{\cal N}_{\rm HI}}(\nu_{\rm i}) &=& 4\pi f_{\rm sky}\frac{T^2_{\rm sys}(\nu_{\rm i})}{2N_{\rm dish}t_{\rm tot}\Delta\nu} \,,\\
T_{\rm sys} &=& 25 + 60\left(\frac{300\,\text{MHz}}{\nu}\right)^{2.55}\ \text{K} \,, 
\ea
where $N_{\rm dish} = 195$, $D_{\rm dish} = 15\,$m, and the system temperature includes a constant 
instrument temperature and a sky component.\\

\noindent{\em SKA2 HI galaxy redshift survey:}\\
The SKA2 survey has not yet been designed, so that the specifications can only be indicative. 
We have used the specifications from the relevant chapter of the SKA Science Book~\cite{Abdalla:2015kra}.
The models for the number counts per redshift per deg$^2$ and the bias of the HI galaxy distribution 
are obtained in~\cite{Yahya:2014yva,Santos:2015dsa} by fitting the simulated data:
\be
\label{eqn:galSKA2}
\frac{\dd N}{\dd z} = 10^{c_1}z^{c_2}e^{-c_3z} \,,\qquad
b_{\rm g} = c_4 e^{c_5z} \,,
\ee
where the coefficients $c_{a}$ depend on the flux limit of the experiment.
For SKA2, we assume a total observation time of $10^5\,$hours, over 30,000\,deg$^2$ 
($f_{\rm sky} \simeq 0.73$) in a redshift range $0.1\le z\le 2$, and with rms flux, constant across the 
band, of $S_{\rm rms}^{\rm ref}=5.4\,\mu$Jy. For this flux, we have~\cite{Yahya:2014yva}:
\be 
c_1=6.555\,,~ c_2=1.932\,,~ c_3=6.378\,, ~ c_4=0.549\,, ~ c_5=0.812 \,.
\ee

\noindent{\em SKA1 continuum survey:}\\
SKA1-MID should detect radio sources out to $z \sim 5$ over 25,000\,deg$^2$ ($f_{\rm sky} \simeq 0.60$), 
with an rms $\sim 1\,\mu$Jy and a source detection limit 5\,$\mu$Jy.
The redshift distribution and bias are predicted by simulations, for each type of radio galaxy. 
These are then combined to produce the total quantities:
\be
N(z)=\sum_a N_a(z)\,,~~b(z) = \sum_a b_a(z)\,{N_a(z)\over N(z)} \,.
\ee 
Details are given in~\cite{Alonso:2015uua}. The continuum survey has a single redshift bin if we do not 
have redshift information from cross-matching with other surveys. We will assume that the survey can be 
split into 5 bins.\\

We summarize the linear bias and the number density of galaxies as functions of redshift for 
the four surveys in figure~\ref{fig:spec}.\\

\noindent{\em Planck:}\\
We consider the {\em Planck} Fisher matrix in combination with the large-scale structure Fisher 
matrices, following the method of~\cite{Ballardini:2016hpi}.
We assume a white noise corresponding to the {\em Planck} 143\,GHz channel updated full mission 
sensitivities of 33\,$\mu$\,arcmin in temperature and 70.2\,$\mu$K\,arcmin in polarization, 
and a beam resolution of 7.3\,arcmin over 29,000\,deg$^2$ ($f_{\rm sky} \simeq 0.70$).

\section{Models of features in the primordial power spectrum}
\label{sec:three}
We consider three inflationary models that generate features in the primordial power spectrum, 
which we will call the {\bf kink} \cite{Starobinsky:1992ts}, {\bf step} \cite{Dvorkin:2009ne} 
and {\bf warp} \cite{Miranda:2012rm} models.
Following~\cite{Ballardini:2016hpi}, we adopt the best fit parameters from {\em Planck} TT+lowP 
\cite{Ade:2015lrj} for the three parametrized models. 

\begin{figure}[h!]
\centering
\includegraphics[width=10cm]{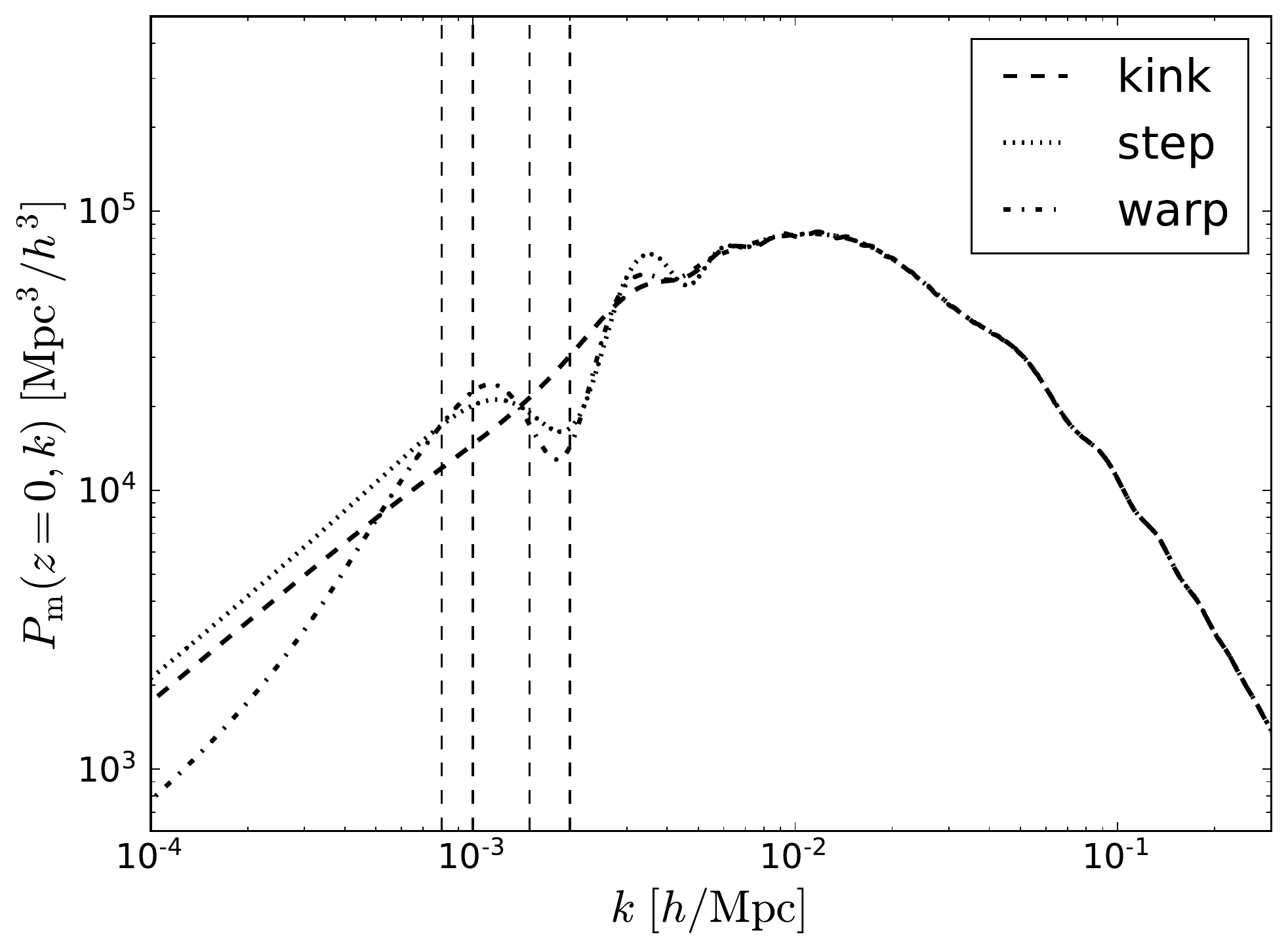}
\caption{
The best fit from {\em Planck} 2015 for the matter power spectrum, for the three primordial feature 
models. Dashed vertical lines correspond to the contour lines in figure~\ref{fig:kmin}.}
\label{fig:Pk}
\end{figure}

\begin{figure}[h!]
\centering
\includegraphics[width=10cm]{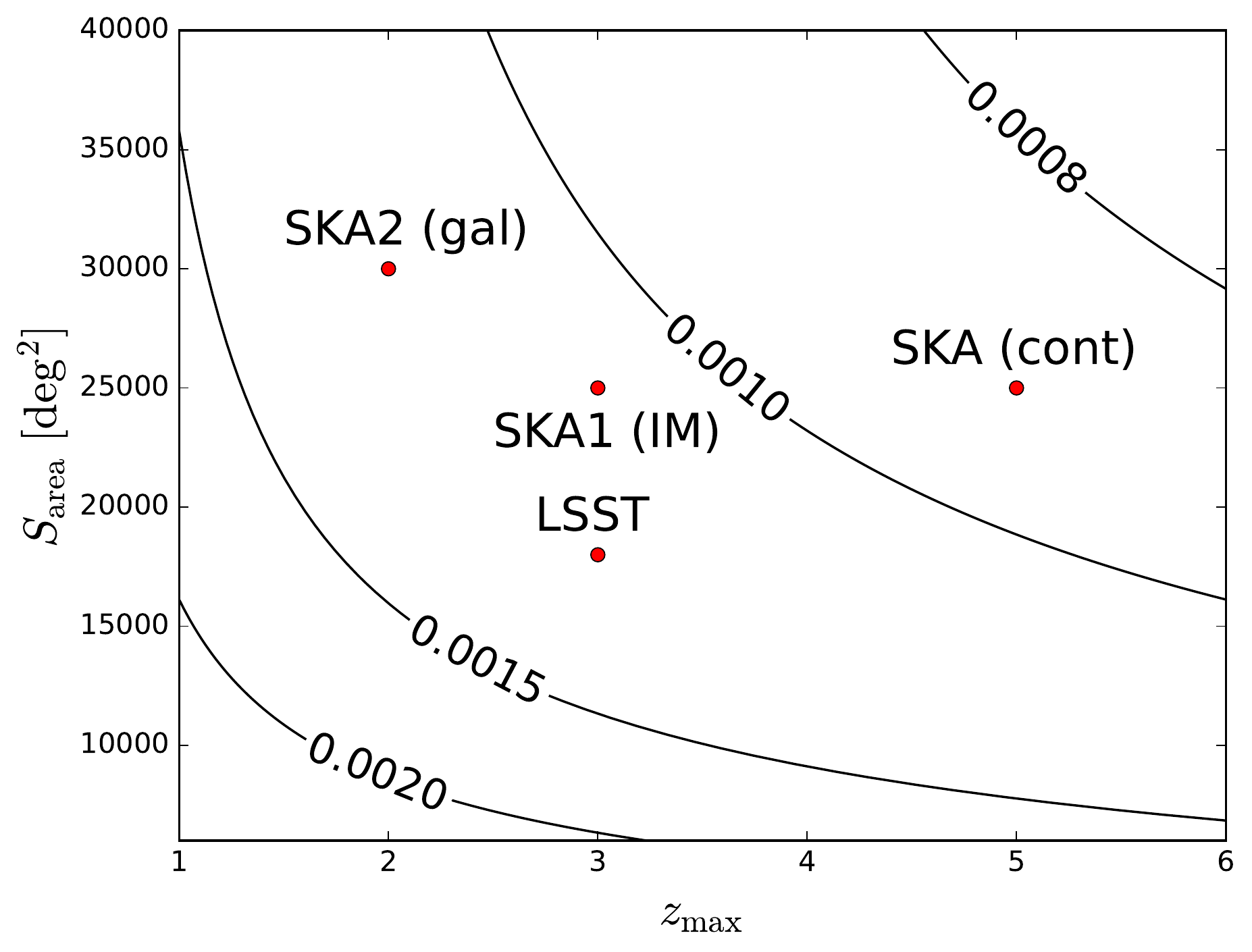}
\caption{The sky area and maximum redshift of the surveys considered. Contours for 
$k_\mathrm{min} = {0.0008}, 0.001, 0.0015, 0.002\,h/$Mpc indicate the largest accessible scale 
estimated to be contained in the survey volume.}
\label{fig:kmin}
\end{figure}

\begin{figure}[h!]
\centering
\includegraphics[width=7.7cm]{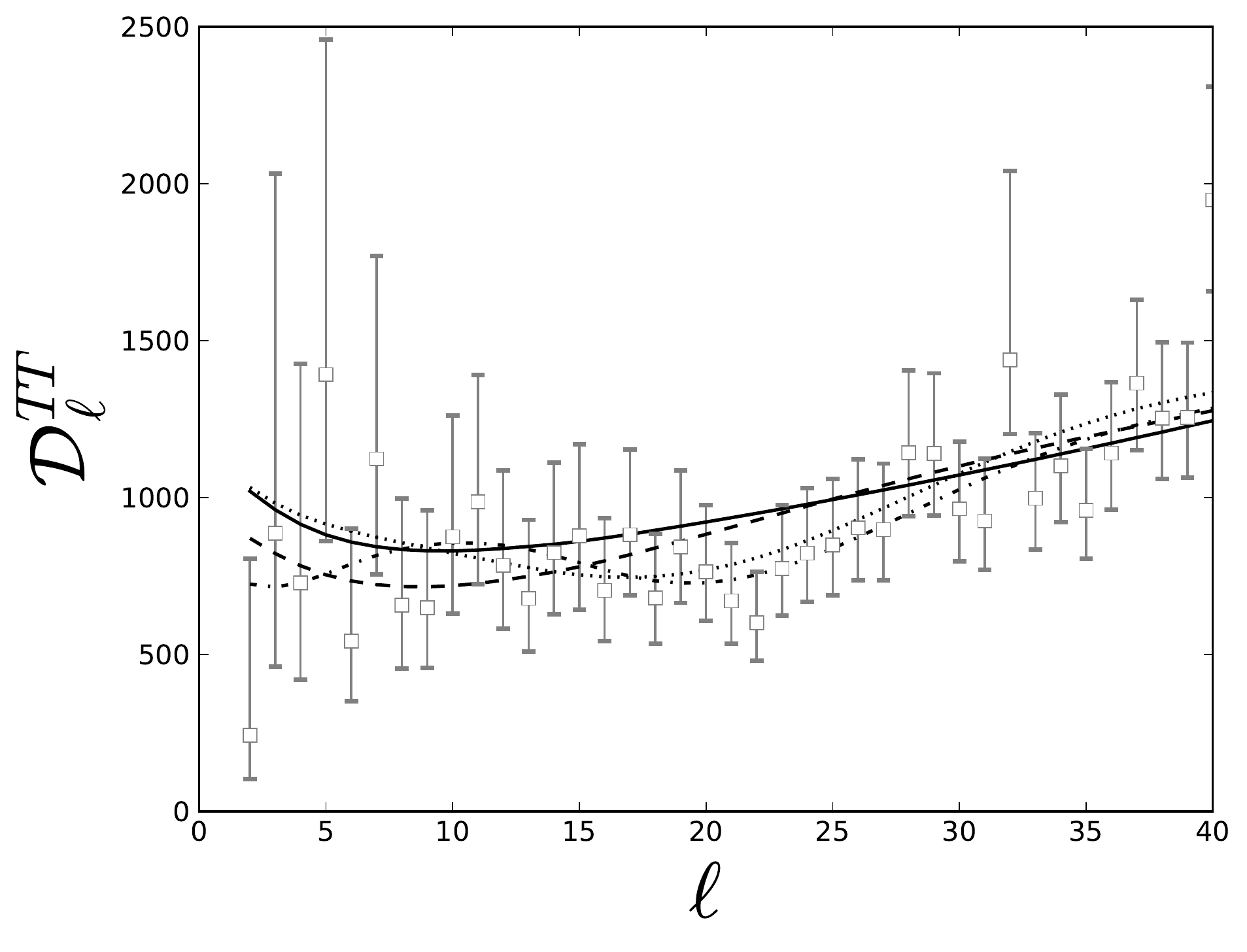}~~
\includegraphics[width=7.7cm]{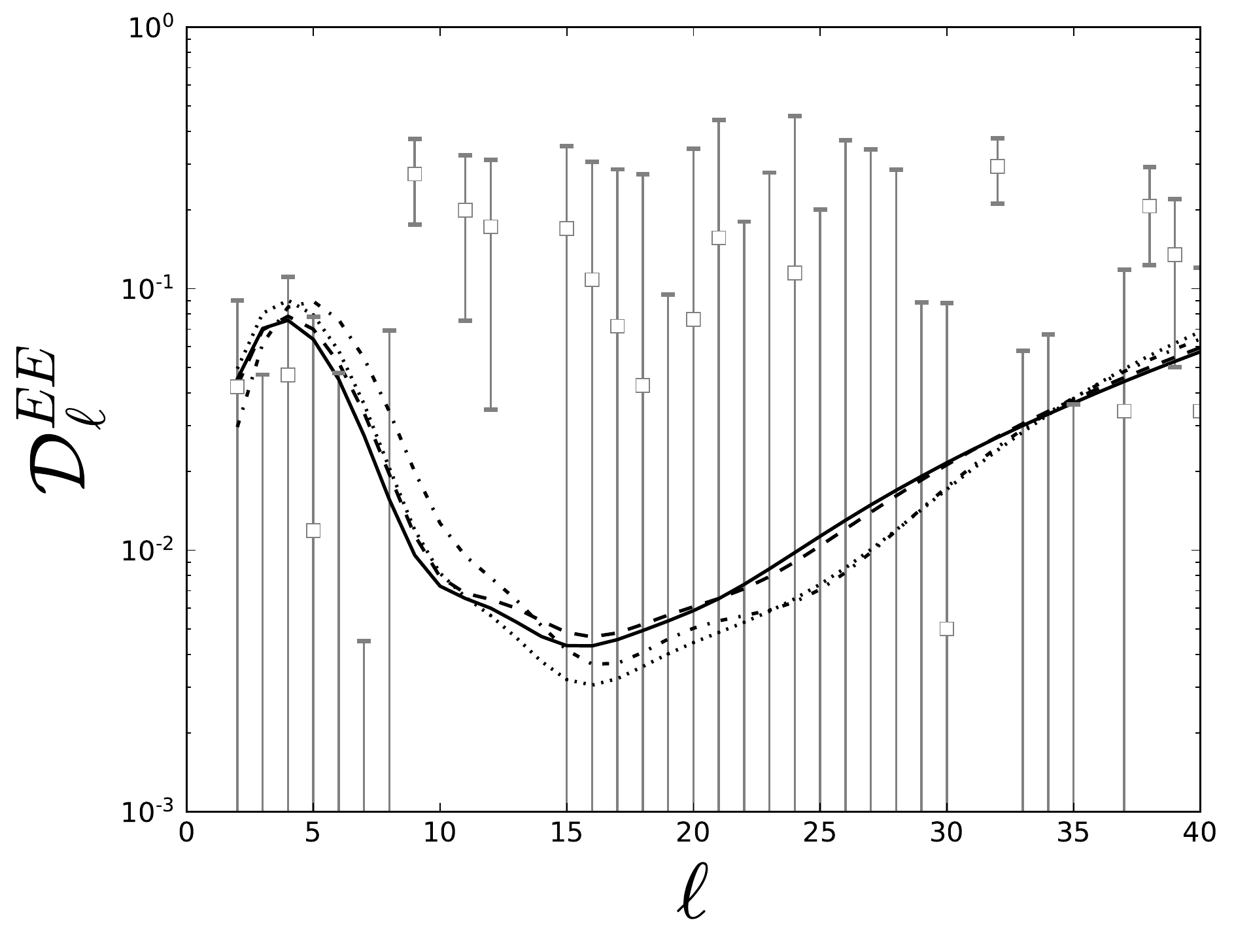}
\caption{The left (right) panel shows the {\em Planck} 2015 temperature (E-mode polarization) 
{data} compared to the best fit spectra for $\Lambda$CDM (solid black line) and the three 
primordial feature models (broken lines -- see Fig.~\ref{fig:Pk}).}
\label{fig:CMB}
\end{figure}

Figure~\ref{fig:Pk} shows the best fit-matter power spectra for the three models, and 
figure~\ref{fig:kmin} indicates some characteristic scales of the models in the sky area/ maximum 
redshift plane. The largest-scale contours are calculated from the comoving volume:
\be
\label{eqn:kmin}
V_{\rm surv}={4\pi\over3}f_{\rm sky}\big[ r(z_{\rm max})^3-r(z_{\rm min})^3\big],\qquad k_\mathrm{min} ={2\pi\over V_{\rm surv}^{1/3}} \,.
\ee
Figure~\ref{fig:CMB} compares the $\Lambda$CDM {\em Planck} 2015  TT and EE power spectra with those 
of the primordial feature models.

We write the primordial power spectrum as the standard featureless  one $\mathcal{P}_{\mathcal{R},0}$, 
modulated by the contribution due to the violation of slow-roll:
\be
\label{eqn:generalPPS}
\mathcal{P}_{\mathcal{R}} (k) = \mathcal{P}_{\mathcal{R},0} (k) \cdot \mathcal{P}_{\mathcal{R},1}(k)\,,\qquad \mathcal{P}_{\mathcal{R},0} (k) = A_{\rm s}\left({k\over k_*}\right)^{n_{\rm s}-1} \,.
\ee

\subsection{Kink model}
This model has a sharp change in the slope of 
the inflaton potential, which is constant near  the transition~\cite{Starobinsky:1992ts}.
After the transition the second slow-roll parameter $\epsilon_2$ becomes large for some time 
because of the discontinuity in the first derivative of the potential -- afterwards, 
slow-roll is recovered.
The two different slopes of the potential lead to different asymptotic values of the curvature 
power spectrum, plus an oscillatory pattern in between. 
The contribution to $\mathcal{P}_{\mathcal{R}}$ can be derived analytically: 
\begin{align}
\mathcal{P}_{\mathcal{R},1}(y) ={} & 1 + \frac{9}{2}\mathcal{A}_\mathrm{kink}^2\left( \frac{1}{y} + \frac{1}{y^3} \right)^2 
+ \frac{3}{2}\mathcal{A}_\mathrm{kink}\left( 4 + 3\mathcal{A}_\mathrm{kink} - 3\frac{\mathcal{A}_\mathrm{kink}}{y^4} \right)^2 \frac{1}{y^2} \cos(2y) \notag\\
&{}+ 3\mathcal{A}_\mathrm{kink}\left( 1 - \frac{1 + 3\mathcal{A}_\mathrm{kink}}{y^2} - \frac{3\mathcal{A}_\mathrm{kink}}{y^4} \right)^2 \frac{1}{y} \sin(2y)\,, \quad y\equiv {k\over k_\mathrm{kink}}\,.
\label{prk}
\end{align}
Here the scale of the transition is $k_\mathrm{kink}$,  the amplitude is $\mathcal{A}_\mathrm{kink}=(A_+-A_-)/A_+$, and we use the approximation 
$|A_+ \phi| \,, |A_- \phi| \ll V_0$.

The best fit for the two extra parameters obtained from {\em Planck} TT+lowP is
$\mathcal{A}_\mathrm{kink} = 0.089$,  
$\log\left(k_\mathrm{kink}\, \text{Mpc}\right) = -3.05$. This provides an improvement 
in the fit of CMB data of $\Delta \chi^2 \simeq -4.5$ \cite{Ade:2015lrj}.

\subsection{Step model}
A step-like feature in the inflaton potential with a discontinuity of the second 
derivative of the potential leads to a localized oscillatory pattern with 
a negligible difference in the asymptotic amplitudes of $\mathcal{P}_{\mathcal{R}}$.
There is an analytical approximation, up to second order in the Green's function 
expansion~\cite{Adams:2001vc,Dvorkin:2009ne,Miranda:2013wxa}:
\ba
\mathcal{P}_{\mathcal{R},1}(y) &=& \exp \Big\{ \mathcal{I}_0(y) + \ln \Big[1+\mathcal{I}_1^2(y)\Big] \Big\} \,,\qquad 
y \equiv {k\over k_\mathrm{step}}\,,
\ea
where $k_\mathrm{step}$ is the inverse of the oscillation frequency. The first- and second-order parts are
\ba
\mathcal{I}_0 (y) &=& \Big[ C_1 W(y) +C_2 W'(y)
+ C_3 Y(y) \Big] \mathcal{D}\left(\frac{y}{ x_{\rm step}}\right)\,,
\label{eqn:step_first_order}\\
\sqrt{2}\,\mathcal{I}_1 (y) &=& \frac{\pi}{2}\left(1-n_{\rm s}\right) 
+ \Big[ C_1 X(y) + C_2 X'(y)
+ C_3 Z(y) \Big] \mathcal{D}\left(\frac{y}{x_{\rm step}}\right)\,,
\label{eqn:step_second_order}
\ea
where $x_\mathrm{step}$ is the damping scale,  a prime denotes ${\rm d}/{\rm d} \ln y$, and 
the damping envelope is:
\be
\label{eqn:damping}
\mathcal{D}(y) = \frac{y}{\sinh y} \,.
\ee
The window functions are:
\begin{align}
W(y) &= \frac{3\sin(2y)}{2y^3} - \frac{3\cos(2y)}{y^2} - \frac{3\sin(2y)}{2y} \,,\\
X(y) &= \frac{3}{y^3} \left(\sin y - y\cos y\right)^2  \,,\\
Y(y) &= \frac{6y\cos(2y)+(4y^2-3)\sin(2y)}{y^3} \,,\\
Z(y) &= -\frac{3+2y^2-(3-4y^2)\cos(2y)-6y\sin(2y)}{y^3} \,.
\end{align}
The step model has 3 parameters:
\be
 C_1=C_3=0\,,~ ~ C_2 = \mathcal{A}_{\rm step}
\,,~~ k_{\rm step}\,,~~x_{\rm step} \,,
\ee
and its $\Delta \chi^2$ is $\simeq -8.6$ for {\em Planck} TT+lowP, corresponding to the best fit 
parameters $\mathcal{A}_{\rm step} = 0.374$, $\log\left(k_{\rm step}\,\text{Mpc}\right) = -3.1$, and 
$\ln x_{\rm step}=0.342$ \cite{Ade:2015lrj}.

\subsection{Warp model}
In DBI models, a step in the potential affects also the kinetic term of the Lagrangian, leading 
to additional signatures in $\mathcal{P}_{\mathcal{R}}$ -- this is the warp model~\cite{Miranda:2012rm}.
This extention of the step model has 5 parameters: all three $C_a$ in 
\eqref{eqn:step_first_order}--\eqref{eqn:step_second_order} are nonzero.
For the warp model, the $\Delta \chi^2$ increases to $-12.1$ for {\em Planck} TT+lowP, given the cost 
of adding five extra parameters~\cite{Ade:2015lrj}.
The best fit to {\em Planck} TT+lowP is $C_1 = -1.05$, $C_2= \mathcal{A}_{\rm warp} = 1.16$, 
$C_3 = -0.737$, $\log\left(k_{\rm step}\ \text{Mpc}\right) = -3.12$, and $\ln x_{\rm step}=-0.195$.

\section{Results}
\label{sec:four}
We forecast the constraints on the $\mathcal{P}_{\mathcal{R}}$ parameters above, for the LSST and 
SKA surveys (with and without {\em Planck} data), in order to quantify the possibility of 
discriminating these models from a standard power law. (In the limit of a zero amplitude, we 
recover standard power law predictions for all 3 models.)
In the appendix, we give the constraints on the standard cosmological parameters 
with and without the inclusion of the CMB and show the contours together with the constraints on the 
primordial-feature parameters from the CMB alone as comparison.

\begin{figure}[h!]
\centering
\includegraphics[width=7.7cm]{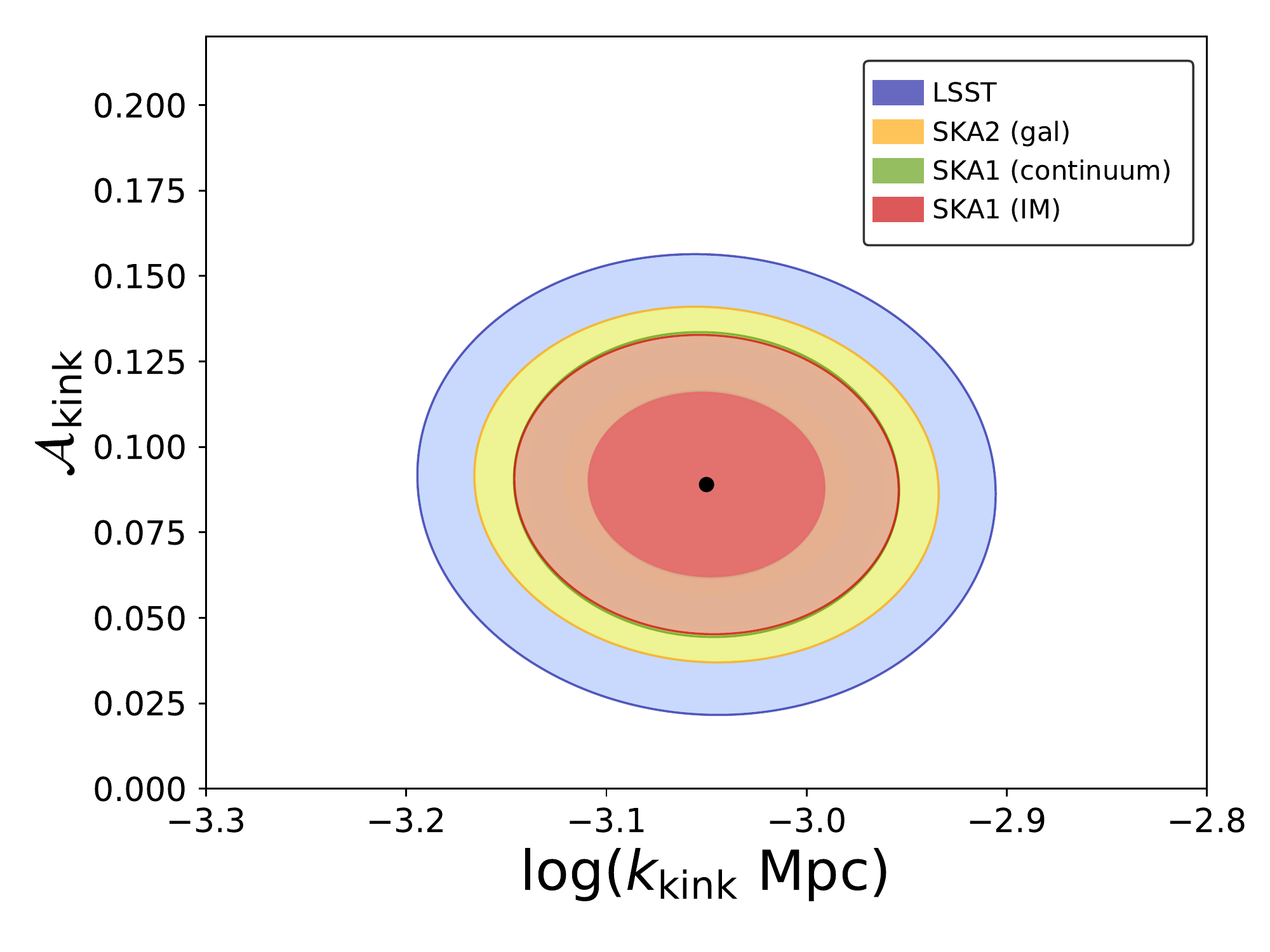}~~
\includegraphics[width=7.7cm]{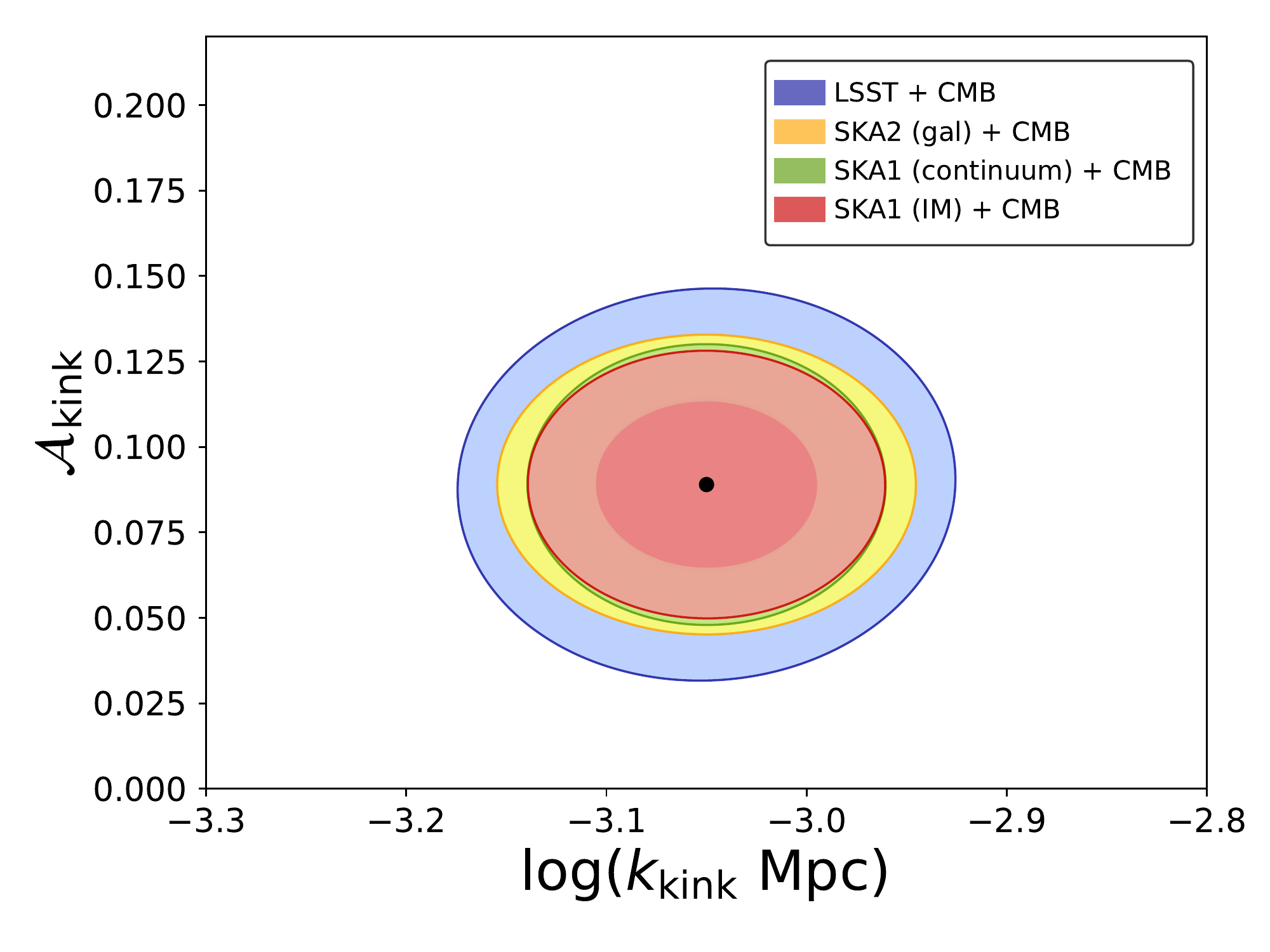}
\caption{Marginalized 68\% and 95\% CL contours for the {\bf kink} model parameters, 
$k_\mathrm{kink}$ and $\mathcal{A}_\mathrm{kink}$, using the surveys alone {\em (left)}, 
and combining survey and CMB Fisher information {\em (right)}.}
\label{fig:2D_kink}
\end{figure}

\begin{figure}[h!]
\centering
\includegraphics[width=7.7cm]{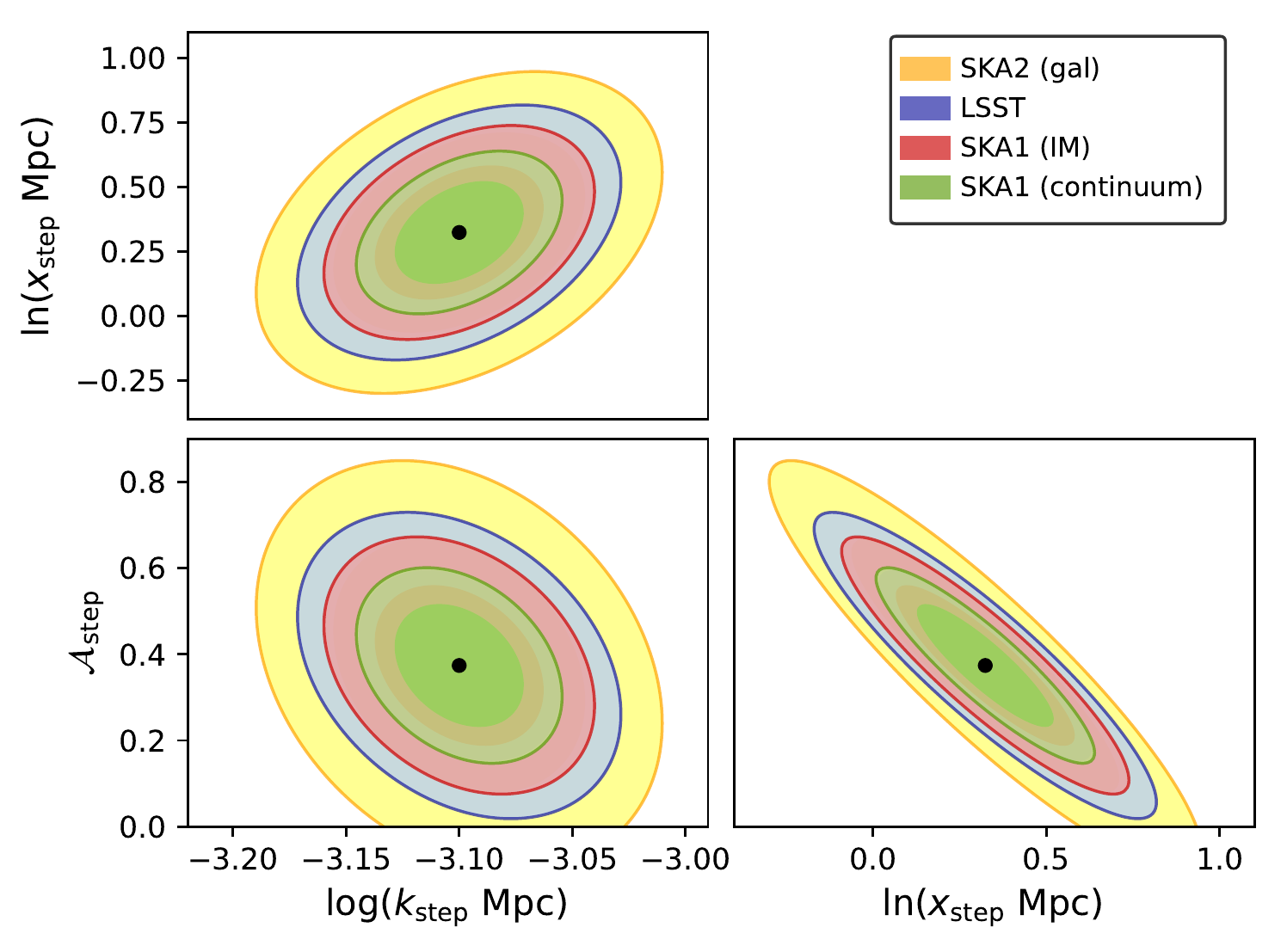}~~
\includegraphics[width=7.7cm]{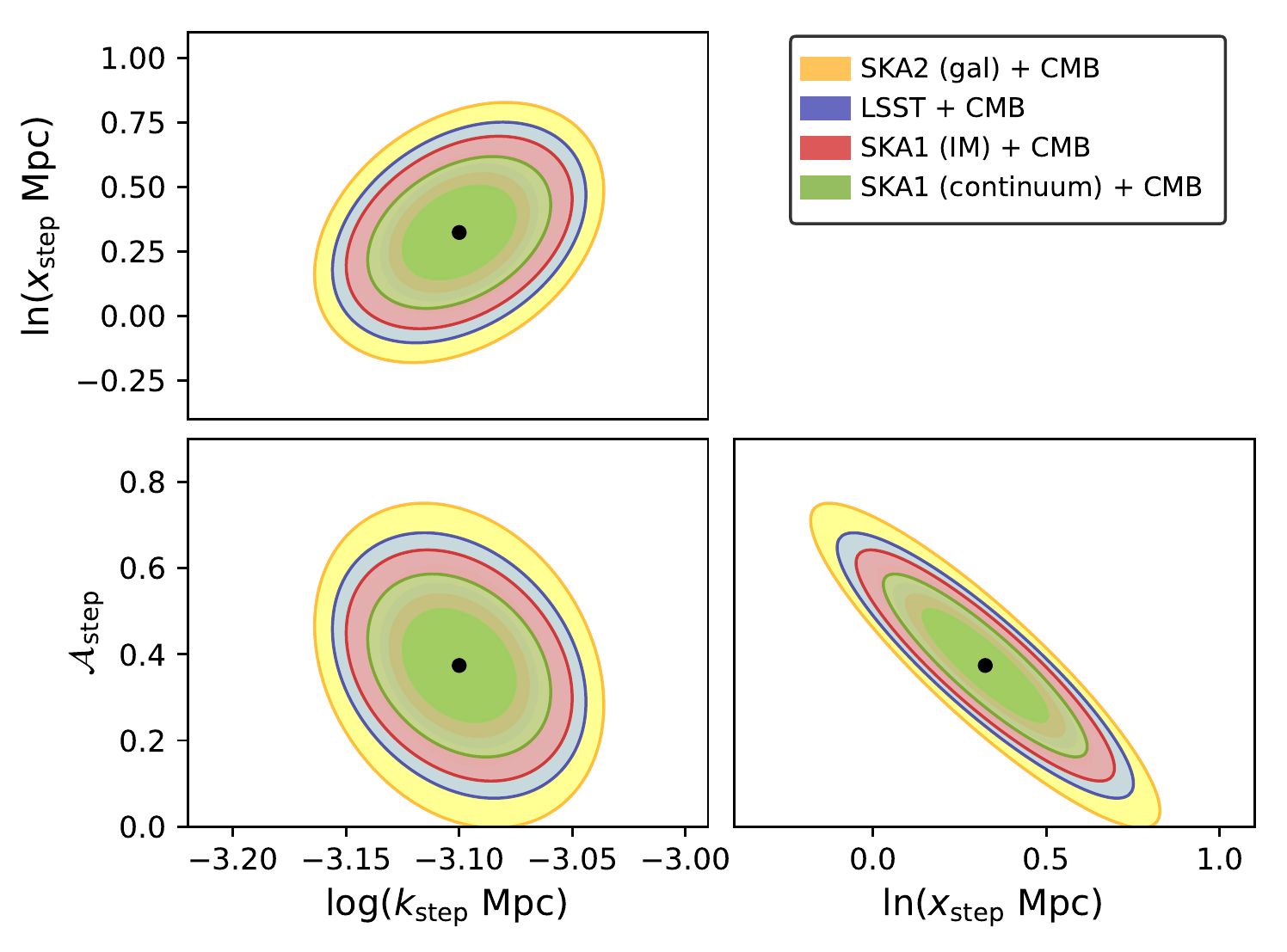}
\caption{As in figure~\ref{fig:2D_kink}, for the parameters of the {\bf step} model.}
\label{fig:2D_step}
\end{figure}

\begin{figure}[h!]
\centering
\includegraphics[width=12cm]{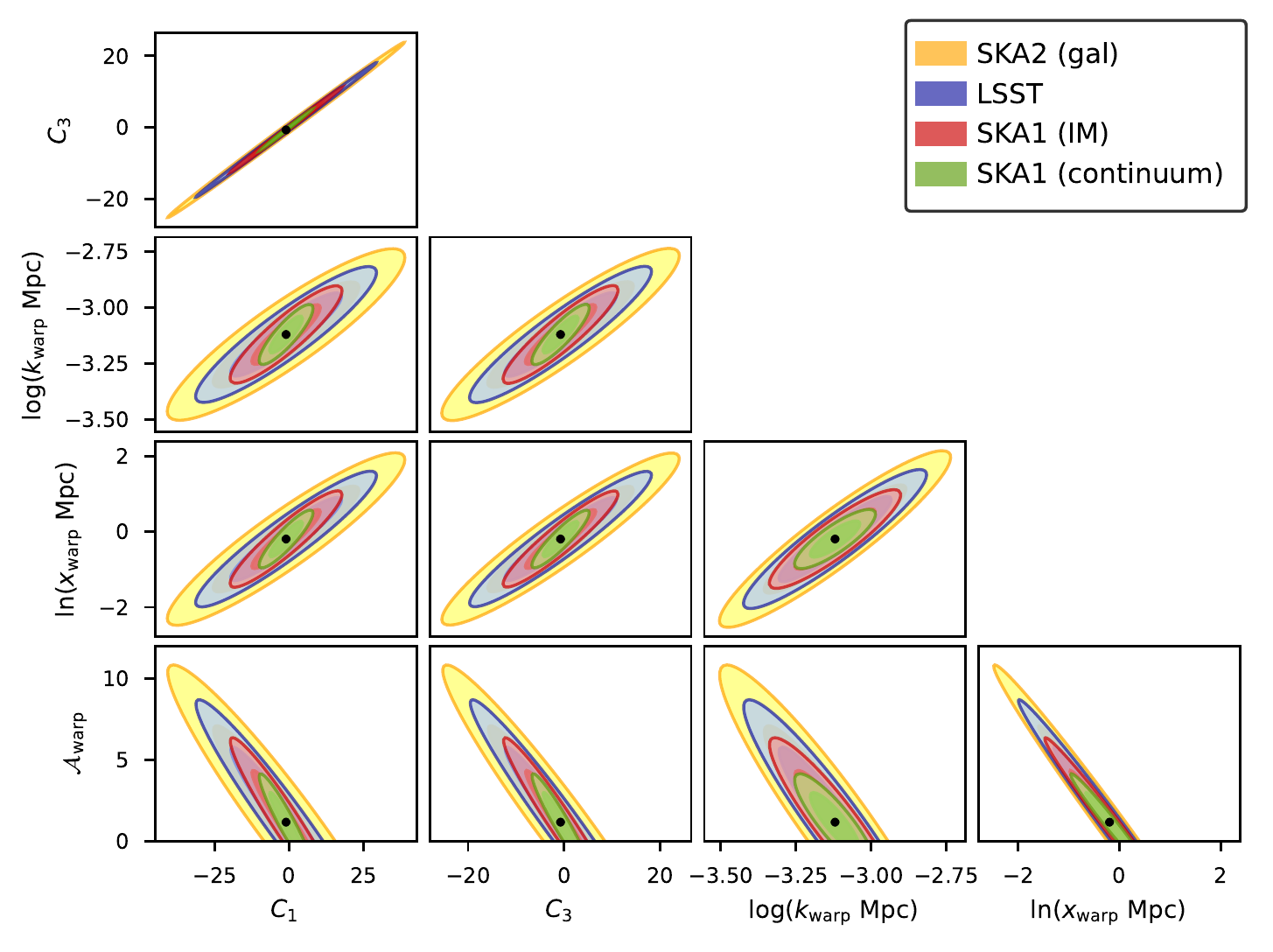}\\
\includegraphics[width=12cm]{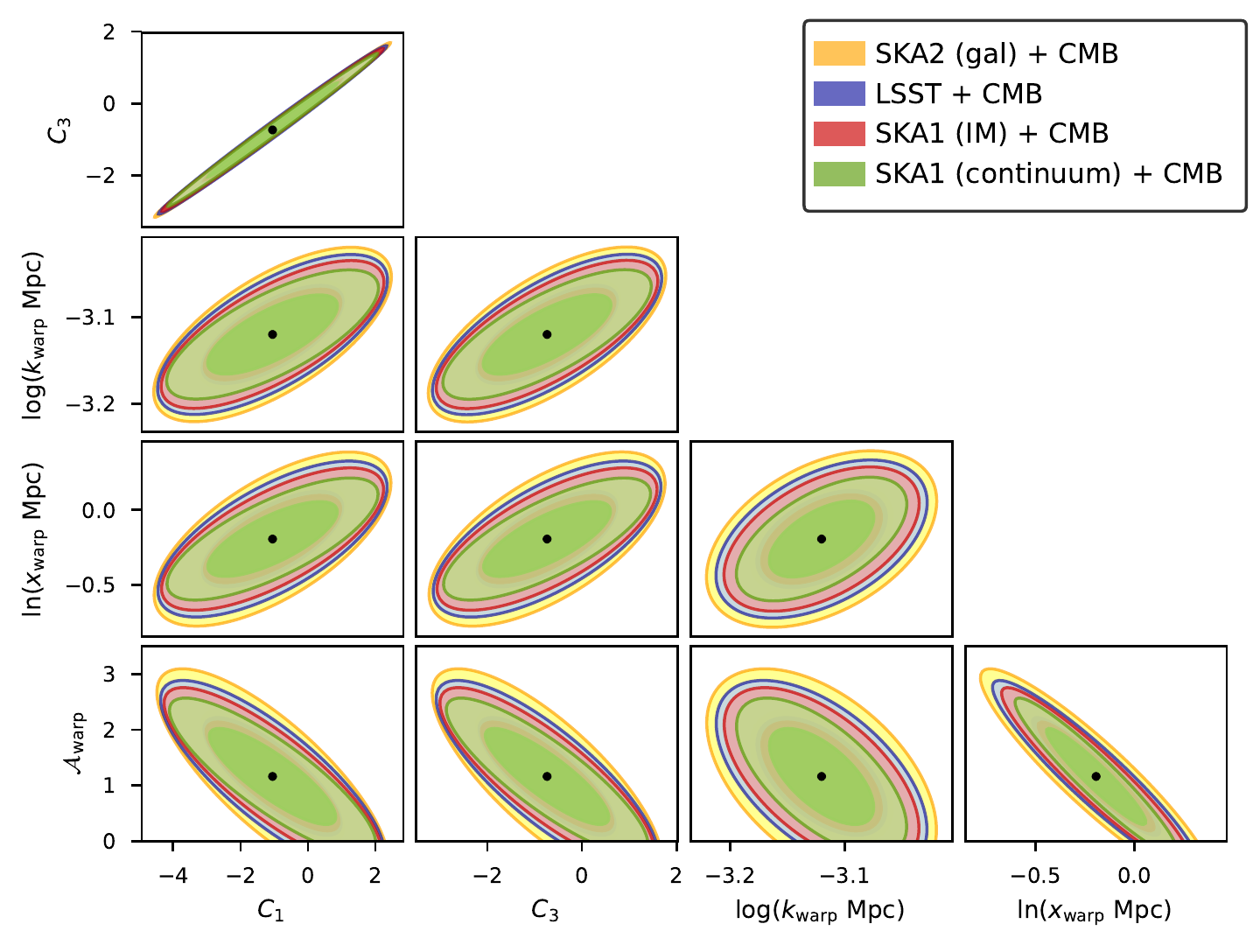}
\caption{As in figures~\ref{fig:2D_kink} and \ref{fig:2D_step}, for the parameters of the {\bf warp} 
model.}
\label{fig:2D_warp}
\end{figure}

We use 5 redshift bins for all the surveys. We have checked that increasing the number of bins 
to $\sim10$ or even $\sim20$ does not change the constraints by more than $1\sigma$, in particular 
when CMB is added.
For LSST and SKA1 IM, we consider the same 5 redshift bins  since they cover the same redshift 
range -- with redshift edges 0.15, 0.5, 1, 1.5, 2.2, 3. These bins have approximately the same comoving 
radial extent. For the SKA2 HI galaxy redshift survey, we use different edges because of the different 
redshift range considered for this survey.
For the  SKA1 continuum survey we assume 5 bins with edges 0, 0.5, 1, 1.5, 2, 5.

Measurement of the power spectrum is affected by the $k$-space window function which depends on the 
limited survey volume observed. We used a top-hat window function. 
We checked that with a Gaussian window function as in~\cite{Huang:2012mr,Chen:2016vvw}, the constraints 
are changed by less than 5\%, thanks to the huge volume covered by these surveys.

Figure~\ref{fig:2D_kink} shows the forecast constraints for the {\bf kink} model.
For this model there are already strong constraints from {\em Planck} 2015 data \cite{Ade:2015lrj}, 
mainly thanks to the oscillatory pattern which leaves an imprint on intermediate scales.
\footnote{The width of the oscillatory pattern is fixed because for this model the transition between 
the two different slopes of the inflaton potential is assumed to be instantaneous. Extensions for a 
non-instantaneous transition have been considered for the bispectrum in~\cite{Martin:2014kja}, at the 
cost of adding an extra parameter to smooth the transition.} 
The best fit from CMB data suggests small deviations from the standard power law prediction, 
which correspond to 20\% at $k\simeq 0.001\, h$/Mpc and less than 10\% at $k\simeq 0.01\, h$/Mpc. 
The results obtained in~\cite{Ballardini:2016hpi}, using spectroscopic surveys with smaller volume 
than LSST and SKA, showed a significant improvement in the constraints on the extra parameters 
with respect to CMB data, thanks to the possibility of recovering the oscillatory pattern in the 3D 
matter power spectrum. In~\cite{Ballardini:2016hpi} it was found that the amplitude could be 
constrained at more than $1\sigma$. 
By using larger-volume surveys, Fig.~\ref{fig:2D_kink} shows that this model can be distinguished from 
the simplest power-law spectrum at more than $3\sigma$, even without CMB data.

We show in figure~\ref{fig:2D_step} the forecast constraints for the parameters of the {\bf step} model. 
This model with a localized feature leads to bigger deviations from standard slow-roll predictions, up 
to 40\% in the range $0.002\, h/{\rm Mpc}  < k < 0.006\, h/{\rm Mpc}$.
Previous studies~\cite{Ballardini:2016hpi,Chen:2016vvw} found that the combination of optical galaxy 
surveys (such as Euclid and LSST) and CMB data will not significantly improve constraints, because the 
feature is located at scales where the galaxy signal to noise is not optimal. Thanks to the SKA it will 
be possible to test the current {\em Planck} 2015 best fit for this model at more than $3\sigma$, as 
shown by figure~\ref{fig:2D_step}. We find that the $1\sigma$ constraint on the amplitude goes from 0.2 
for LSST to 0.009 for the SKA1 radio continuum survey.

Figure~\ref{fig:2D_warp} shows the forecast constraints for the parameters of the {\bf warp} model. 
The CMB best fit for this model shows more pronounced features around $k \sim 0.002\, h/$Mpc.
The relative difference with respect to the slow-roll predictions of the $\Lambda$CDM model is larger 
for this model than for the previous two models, but the degeneracy among the five extra parameters 
makes it difficult to test its predictions with large-scale structure alone.
Another important difference is that for this model the oscillations on intermediate scales are 
smaller, almost absent, and the extent of large-scale suppression in this model, relegated to very 
large scales $k < 0.0005\, h/$Mpc, is not constrained.
The degeneracy among the extra parameters of this model is partially broken when the CMB information 
is added. We find that the combination of large-scale structure and CMB data is really promising, in 
particular to constrain more complex models such as this one.

\subsection{The impact of systematics on the largest scales}
One of the most important challenges in observing on ultra-large scales in the matter distribution 
is the presence of foreground and systematic contamination; see for example different studies on 
systematic uncertainties in BOSS data~\cite{Ross:2011cz,Ho:2012vy,Hernandez-Monteagudo:2013vwa}.
Here we make a rough estimate of the effect of foreground and systematic contamination as follows:
we mimic the effect of losing information on the largest scales by increasing
$k_{\rm min}$ in \eqref{eqn:kmin} by a factor of 2, 3, 4, and 5.
\begin{figure}[h!]
\centering
\includegraphics[width=14cm]{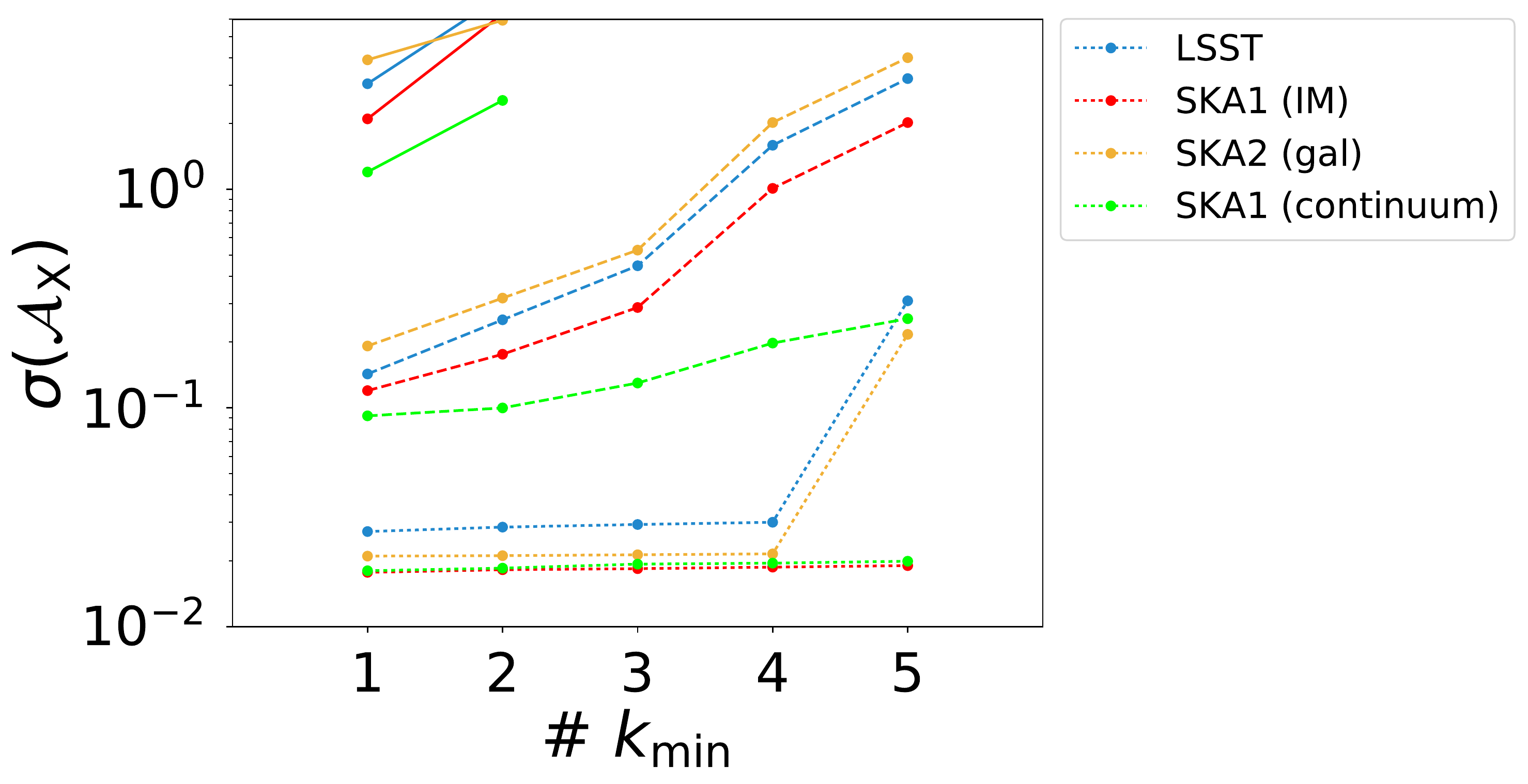}
\caption{Errors on the amplitude ${\cal A}_{\rm X}$ for $X=$ {\bf kink}, {\bf step}, and {\bf warp} 
models, from bottom to top respectively, when the theoretical $k_{\rm min}$ is increased by factors 
of 2,3,4,5. For each model, the errors from the 4 surveys are shown:
LSST (blue), SKA1 IM (red), SKA2 galaxy (yellow), and SKA1 continuum (green).}
\label{fig:kmin_syst}
\end{figure}
The results in figure~\ref{fig:kmin_syst} show that for the {\bf kink} model, the constraints on the
amplitude of the feature are quite robust when $k_{\rm min}$ is increased. We find a bigger impact
on the {\bf step} model, in particular when $k_{\rm min}$ is more than three times larger than
the prediction of \eqref{eqn:kmin}. Finally, for the {\bf warp} model the amplitude of the
feature is unconstrained if we lose the scales $k\lesssim 2k_{\rm min}$.

\subsection{The impact of scale-dependent bias}
Models with deviations from the standard slow-roll inflation generate specific 
shapes in the bispectrum (see \cite{Chen:2010xka} for a review), so that primordial features 
can also be searched for in the bispectrum~\cite{Ade:2015ava}, or jointly in the power spectrum and 
bispectrum~\cite{Fergusson:2014hya,Fergusson:2014tza,Meerburg:2015owa}.
We investigate the additional effect of a scale-dependent bias in the galaxy power spectrum 
induced by the presence of primordial non-Gaussianity from the primordial features.

We use the analytic expressions for the scale-dependent bias from the integrated perturbation theory 
formalism~\cite{Matsubara:2012nc}:
\ba
\Delta b(k) &\simeq& \frac{\sigma_{\rm M}^2}{2\delta_\textrm{c}}
\left[A_2\, \mathcal{I}(k) + A_1 \frac{\partial \mathcal{I}(k)}{\partial \ln \sigma_\textrm{M}} \right] \,,
\\ \label{eqn:Ik2}
\mathcal{I}(k) &\simeq& \frac{1}{\sigma_\mathrm{M}^2 P(k)}\int 
\frac{d^3 k'}{(2\pi)^3}W^2(k'R)B_\textrm{L}(k,k',|\mathbf{k}-\mathbf{k'}|),
\ea
where $A_1$ and $A_2$ have been choosen according to~\cite{Matsubara:2012nc}.
The matter bispectrum is given by
\be
B_\mathrm{L}(k_1,k_2,k_3) = \mathcal{M}(k_1)\mathcal{M}(k_2)\mathcal{M}(k_3)B_\Phi(k_1,k_2,k_3),~~\delta_\mathrm{m}(k)=\mathcal{M}(k)\Phi(k)\,.
\ee
For the step model, we considered the analytical template, up to the second order in the Green's 
function expansion, derived in~\cite{Adshead:2011jq}. 

For the three primordial feature models, the equilateral configuration is the dominant one, as 
shown by~\cite{Gong:2017yih}. 
Hence the scale-dependent bias is not enhanced for this class of models as it would be for a 
local-type primordial non-Gaussianity. 
By including a scale-dependent bias due to the primordial non-Gaussian signal sourced by 
only the violation of slow-roll, we find a small improvement on the constraints, around 3\%, on 
both the amplitude and the position of the feature.

\section{Conclusions}
\label{sec:conclusion}
In this paper we investigated how well some future LSS surveys will be able to improve current 
CMB constraints on features in the primordial power spectrum. Such features may be related to 
the large-scale anomalies in the CMB angular power spectra, seen in {\em Planck}~\cite{Ade:2015lrj} 
and previously in WMAP~\cite{Peiris:2003ff} data.
We focused on effects on the largest scales; features on intermediate and smaller scales of the 
primordial power spectrum have been studied by combining CMB with LSS data 
in~\cite{Chen:2016vvw,Chen:2016zuu,Ballardini:2016hpi,Xu:2016kwz,Fard:2017oex,LHuillier:2017lgm}.

We showed that the upcoming photometric survey with LSST and radio surveys with SKA can 
significantly improve the constraints on the extra parameters of models with features localized at 
ultra-large scales of the matter power spectrum, thanks to the huge volumes probed by these surveys. 
We performed forecasts for three different models with parametrized features in the primordial power 
spectrum, namely the {\bf kink}, {\bf step} and {\bf warp} models described in section~\ref{sec:three}, 
using galaxy clustering as our observable.

LSST and SKA are expected to improve the constraints on feature models, as shown in 
figures~\ref{fig:2D_kink}--\ref{fig:2D_step}-\ref{fig:2D_warp}, due to the large area and redshift 
range covered. In particular, upcoming intensity mapping and radio continuum surveys in Phase 1 of 
the SKA will put extremely tight constraints on the parameters of the feature models at more than 
3$\sigma$, without adding CMB information.
\footnote{Euclid will also perform a galaxy photometric survey whose specifications are used for 
weak lensing forecasts. An optimistic use of the Euclid photometric specifications would lead to 
results which would be slightly worst than LSST, mainly due to the smaller redshift volume probed 
which is crucial for the models considered. We get $\sigma({\cal A}_X) = 0.04 (0.03), 0.20 (0.15), 
5.5 (0.3)$ for the {\bf kink} model, the {\bf step} model and the {\bf warp} model respectively 
(in combination with CMB).}
Even if these experiments alone can improve the constraints for such models, the synergies with CMB 
measurements will be crucial to constrain models with many extra parameters, such as the {\bf warp} 
model, as shown in figure~\ref{fig:2D_warp}.

For all three models studied, we find that the SKA1 radio continuum survey gives the tightest 
constraints on the amplitude of the feature, i.e. 
\ba
{\cal A}_{\rm kink} = 0.089 \pm 0.018\,,\quad 
{\cal A}_{\rm step} = 0.374 \pm 0.092\,, \quad {\cal A}_{\rm warp} = 1.16 \pm 1.20\quad \mbox{at 68\% CL.}
\ea
We have assumed sufficient redshift information to divide the continuum survey into 5 redshift bins. 
However, the constraints do not degrade noticeably when 2 bins are used. It appears that volume is 
more important than redshift information.

We considered the impact of a number of factors that could affect the forecasts:
\begin{itemize}
\item Number of redshift bins: increasing the number of bins does not change the constraints by 
more than $1\sigma$.
\item Window function: top-hat or Gaussian choice changes the constraints by less than 5\%. 
\item Systematics on the largest scales: by increasing the largest theoretical scale, corresponding 
to $k_{\rm min}$, we estimated the degradation of constraints due to loss of information. 
Figure~\ref{fig:kmin_syst} shows that constraints on the amplitude in the kink model are little 
affected, those on the step model suffer significantly if we lose scales with $k\lesssim 4k_{\rm min}$, 
and those on the warp model fall away if we lose scales $k\lesssim 2k_{\rm min}$.
\item Scale-dependent bias induced by the non-Gaussianity from primordial features: the inclusion of 
this extra contribution brings a small improvement of a few \% on the constraints. 
\end{itemize}

We have used the galaxy (and intensity) power spectrum in Fourier space, which does not include 
certain effects on correlations, such as lensing magnification and wide-angle correlations. This 
means that we lose some  information that could improve our constraints. Furthermore, there are 
other relativistic effects from observing on the past lightcone which grow on ultra-large scales 
\cite{Yoo:2012se,Camera:2013kpa,Camera:2014bwa,Camera:2014sba,Camera:2015yqa, Raccanelli:2015vla, Alonso:2015uua, Baker:2015bva, Alonso:2015sfa, Fonseca:2015laa, Raccanelli:2015oma, Bull:2015lja, DiDio:2016ykq, Alonso:2016suf, Lorenz:2017iez}. 
These effects could lead to some degeneracies with primordial features and thus weaken the constraints 
at some level. In future work, we will address these two issues by using the galaxy (intensity) 
angular power spectrum $C_\ell^{\rm g}(z,z')$, including lensing and all other relativistic effects, 
as well as wide-angle correlations.

To conclude, we have shown that future photometric and radio surveys can be used to improve our 
knowledge of the primordial Universe and constrain the hint of deviation from the slow-roll inflation 
paradigm pointed from current CMB data.

~\\ \subsection*{Acknowledgments}
MB and RM acknowledge  support from the South African SKA Project. MB is also supported by the Claude 
Leon Foundation, and RM is also supported by the UK STFC (Grant ST/N000668/1).
MB, FF and LM acknowledge the support from the grant MIUR PRIN 2015 ``Cosmology and Fundamental 
Physics: Illuminating the Dark Universe with Euclid'' and from the agreement ASI n.I/023/12/0 
``Attivit\`a relative alla fase B2/C per la missione Euclid''. 
FF wishes to thank UWC for warm hospitality.

\newpage
\appendix
\section{Additional constraints}
\label{sec:appA}

We report the uncertainties in the cosmological parameters at 68\% CL for the kink, 
step and warp models in tables~\ref{tab:errors_kink}, \ref{tab:errors_step} and \ref{tab:errors_warp} 
respectively.

\begin{table}[!h]
\centering
\caption{Constraints on the cosmological parameters for the {\bf kink} model using LSS surveys alone. In parentheses: combining LSS survey and CMB Fisher information.}
\label{tab:errors_kink}
\vspace*{0.2cm}
\begin{tabular}{|c|cccc|}
\hline
\rule[-1mm]{0mm}{.4cm}
 & LSST & SKA2 gal & SKA1 IM & SKA1 cont \\
\hline
\rule[-1mm]{0mm}{.4cm}
$\omega_c$ & 0.0017 (0.00030) & 0.00078 (0.00028) & 0.00087 (0.00028) & 0.0013 (0.00030) \\
\rule[-1mm]{0mm}{.4cm}
$\omega_b$ & 0.00038 (0.000090) & 0.00019 (0.000081) & 0.00021 (0.000082) & 0.00028 (0.000088) \\
\rule[-1mm]{0mm}{.4cm}
$h$ & 0.0028 (0.0015) & 0.0014 (0.00090) & 0.0015 (0.0010) & 0.0020 (0.0012) \\
\rule[-1mm]{0mm}{.4cm}
$n_s$ & 0.0096 (0.0020) & 0.0047 (0.0019) & 0.0052 (0.0018) & 0.0070 (0.0019) \\
\rule[-1mm]{0mm}{.4cm}
$\log\left(10^{10}A_s\right)$ & 0.051 (0.010) & 0.0061 (0.0048) & 0.0055 (0.0042) & 0.037 (0.0098) \\
\rule[-1mm]{0mm}{.4cm}
${\cal A}_{\rm kink}$ & 0.027 (0.023) & 0.022 (0.018) & 0.018 (0.016) & 0.018 (0.016) \\
\rule[-1mm]{0mm}{.4cm}
$\log\left(k_{\rm kink}\, {\rm Mpc}\right)$ & 0.058 (0.050) & 0.050 (0.042) & 0.039 (0.036) & 0.039 (0.036) \\
\hline
\end{tabular}
\end{table}
\begin{table}[!h]
\centering
\caption{As in Table \ref{tab:errors_kink}, for the {\bf step} model.}
\label{tab:errors_step}
\vspace*{0.2cm}
\begin{tabular}{|c|cccc|}
\hline
\rule[-1mm]{0mm}{.4cm}
 & LSST & SKA2 gal & SKA1 IM & SKA1 cont \\
\hline
\rule[-1mm]{0mm}{.4cm}
$\omega_c$ & 0.0014 (0.00028) & 0.00069 (0.00022) & 0.00071 (0.00023) & 0.00099 (0.00026) \\
\rule[-1mm]{0mm}{.4cm}
$\omega_b$ & 0.00027 (0.000083) & 0.00014 (0.000061) & 0.00014 (0.000065) & 0.00020 (0.000076) \\
\rule[-1mm]{0mm}{.4cm}
$h$ & 0.0051 (0.0030) & 0.0028 (0.0022) & 0.0029 (0.0020) & 0.0037 (0.0023) \\
\rule[-1mm]{0mm}{.4cm}
$n_s$ & 0.0095 (0.0022) & 0.0047 (0.0020) & 0.0053 (0.0020) & 0.0069 (0.0021) \\
\rule[-1mm]{0mm}{.4cm}
$\log\left(10^{10}A_s\right)$ & 0.052 (0.0091) & 0.0083 (0.0052) & 0.0072 (0.0053) & 0.038 (0.0089) \\
\rule[-1mm]{0mm}{.4cm}
${\cal A}_{\rm step}$ & 0.14 (0.12) & 0.19 (0.14) & 0.12 (0.11) & 0.092 (0.084) \\
\rule[-1mm]{0mm}{.4cm}
$\log\left(k_{\rm step}\, {\rm Mpc}\right)$ & 0.029 (0.022) & 0.036 (0.025) & 0.024 (0.19) & 0.018 (0.016) \\
\rule[-1mm]{0mm}{.4cm}
$\ln\left(x_{\rm step}\, {\rm Mpc}\right)$ & 0.20 (0.17) & 0.25 (0.19) & 0.17 (0.15) & 0.13 (0.12) \\
\hline
\end{tabular}
\end{table}
\begin{table}
\centering
\caption{As in Table \ref{tab:errors_kink}, for the {\bf warp} model.}
\label{tab:errors_warp}
\vspace*{0.2cm}
\begin{tabular}{|c|cccc|}
\hline
\rule[-1mm]{0mm}{.4cm}
 & LSST & SKA2 gal & SKA1 IM & SKA1 cont \\
\hline
\rule[-1mm]{0mm}{.4cm}
$\omega_c$ & 0.0014 (0.00028) & 0.00069 (0.00022) & 0.00070 (0.00023) & 0.0010 (0.00026) \\
\rule[-1mm]{0mm}{.4cm}
$\omega_b$ & 0.00028 (0.000082) & 0.00014 (0.000061) & 0.00014 (0.000065) & 0.00020 (0.000076) \\
\rule[-1mm]{0mm}{.4cm}
$h$ & 0.0054 (0.0030) & 0.0029 (0.0022) & 0.0029 (0.0021) & 0.0039 (0.0024) \\
\rule[-1mm]{0mm}{.4cm}
$n_s$ & 0.0095 (0.0022) & 0.0047 (0.0020) & 0.0053 (0.0020) & 0.0069 (0.0021) \\
\rule[-1mm]{0mm}{.4cm}
$\log\left(10^{10}A_s\right)$ & 0.052 (0.011) & 0.0083 (0.0055) & 0.0073 (0.0057) & 0.037 (0.011) \\
\rule[-1mm]{0mm}{.4cm}
${\cal A}_{\rm warp}$ & 3.0 (0.70) & 3.9 (0.78) & 2.1 (0.64) & 1.2 (0.35) \\
\rule[-1mm]{0mm}{.4cm}
$\log\left(k_{\rm warp}\, {\rm Mpc}\right)$ & 0.12 (0.037) & 0.15 (0.040) & 0.088 (0.034) & 0.054 (0.027) \\
\rule[-1mm]{0mm}{.4cm}
$\ln\left(x_{\rm warp}\, {\rm Mpc}\right)$ & 0.73 (0.21) & 0.92 (0.23) & 0.52 (0.19) & 0.31 (0.13) \\
\rule[-1mm]{0mm}{.4cm}
$C_1$ & 12.2 (1.34) & 16.0 (1.4) & 7.5 (1.3) & 3.6 (0.45) \\
\rule[-1mm]{0mm}{.4cm}
$C_3$ & 7.6 (0.94) & 9.9 (0.98) & 4.8 (0.91) & 2.4 (0.29) \\
\hline
\end{tabular}
\end{table}

\newpage
\section{Comparison with CMB}
\label{sec:appB}

\begin{figure}[h!]
\centering
\includegraphics[width=7.7cm]{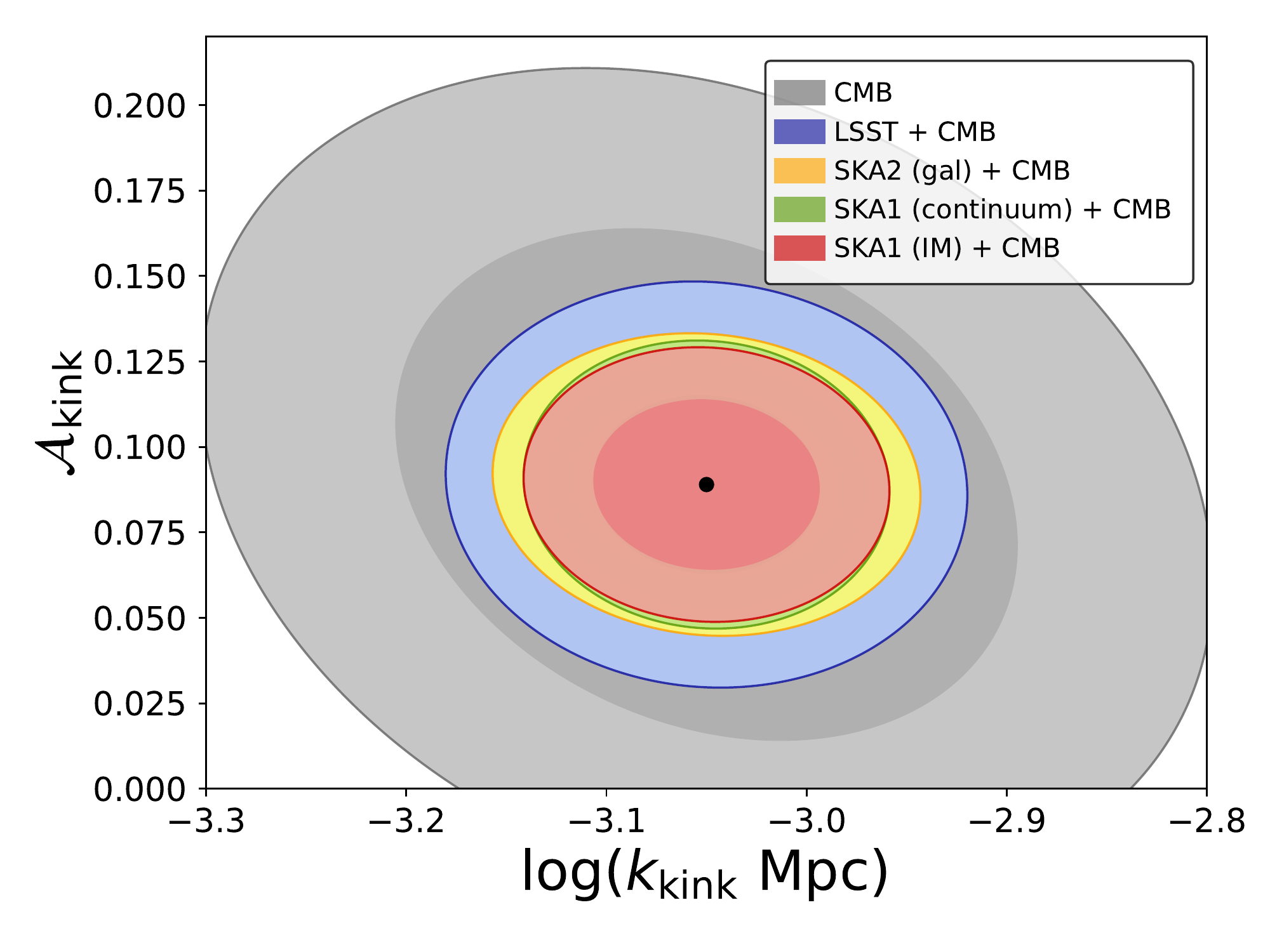}
\caption{As in figure~\ref{fig:2D_kink} plus the constraints from CMB alone.}
\label{fig:kink_CMB}
\end{figure}

\begin{figure}[h!]
\centering
\includegraphics[width=10cm]{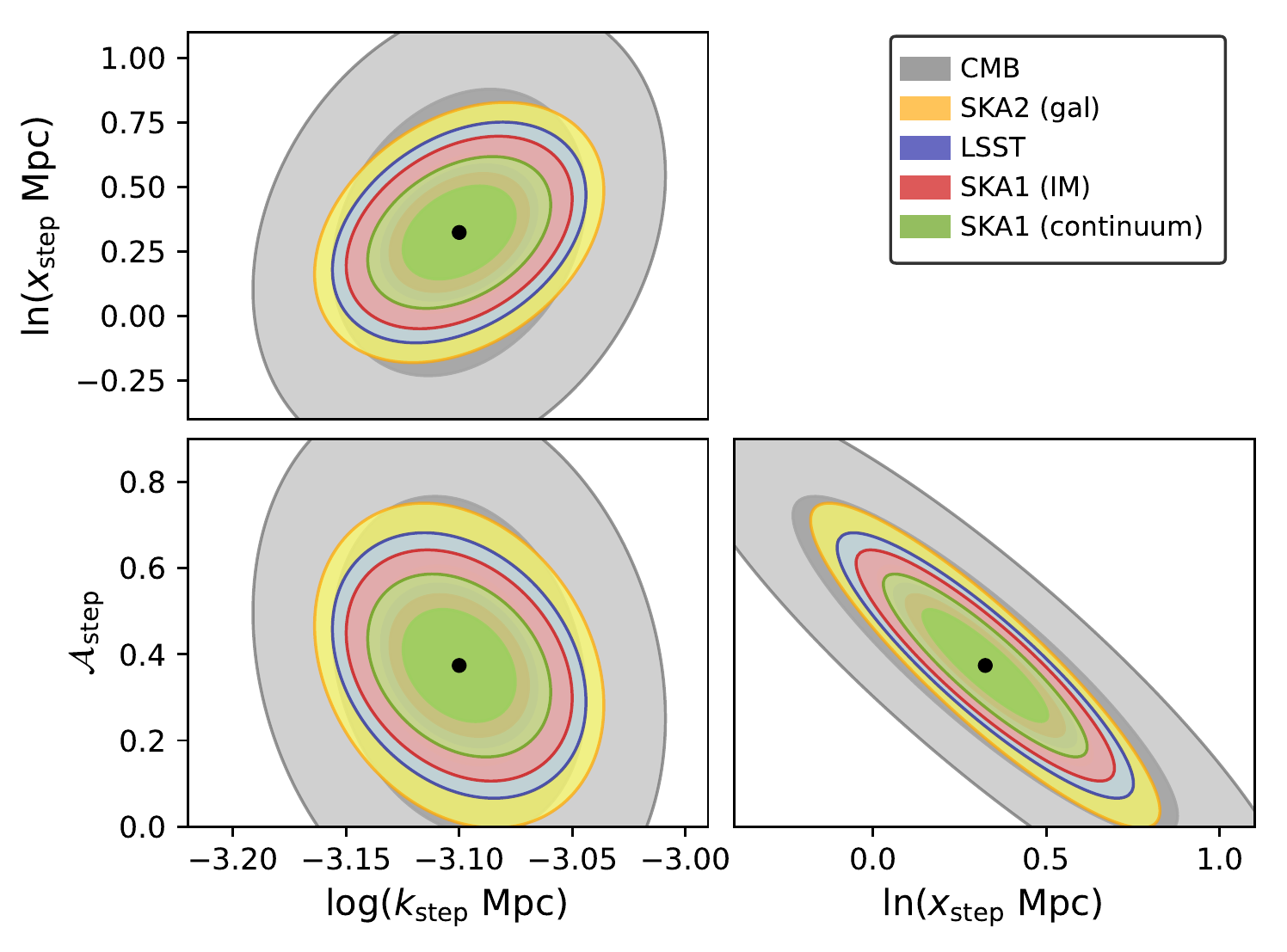}
\caption{As in figure~\ref{fig:2D_step} plus the constraints from CMB alone.}
\label{fig:step_CMB}
\end{figure}

\begin{figure}[h!]
\centering
\includegraphics[width=14cm]{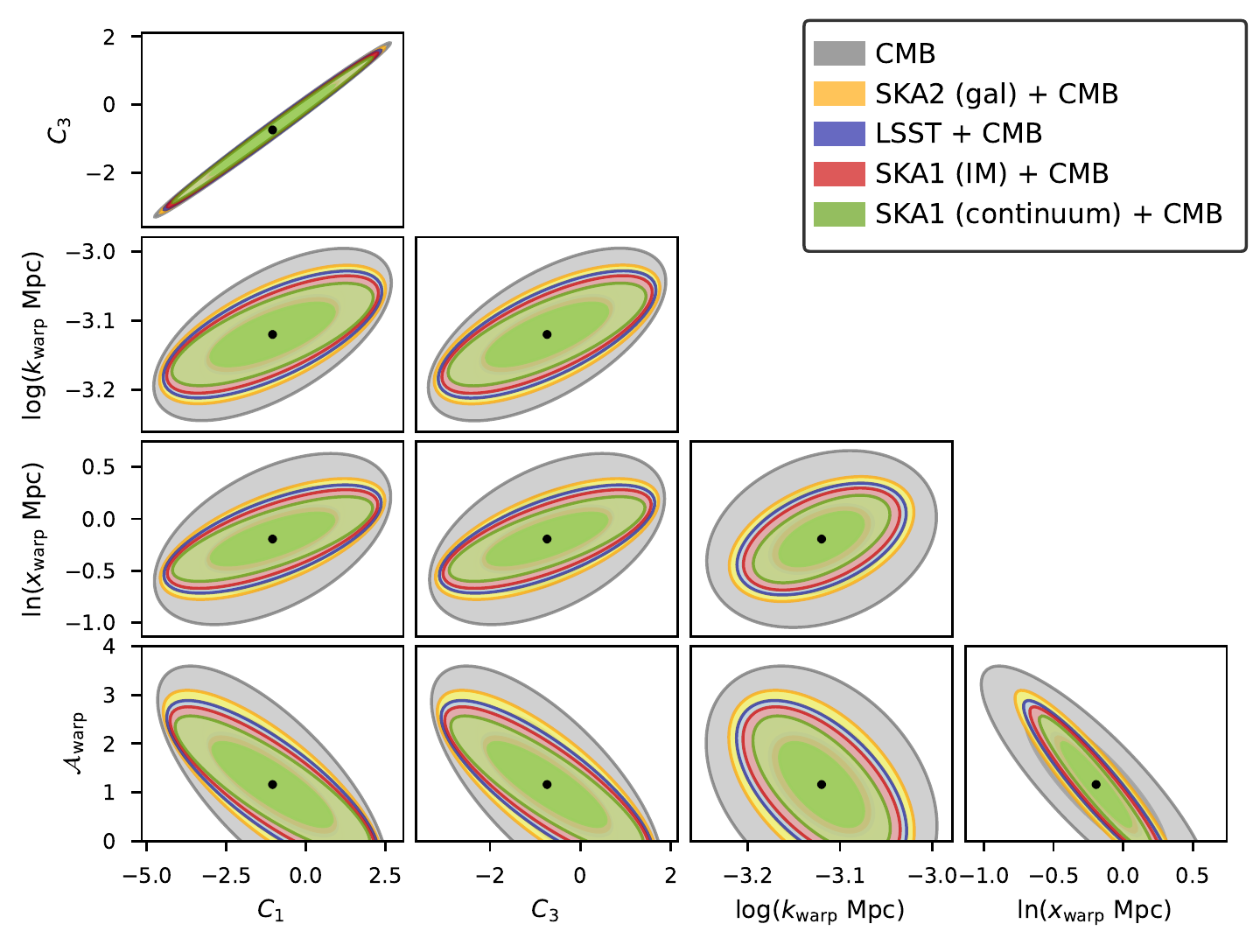}
\caption{As in figure~\ref{fig:2D_warp} plus the constraints from CMB alone.}
\label{fig:warp_CMB}
\end{figure}

We show the marginalized 68\% and 95\%CL for the kink, step and 
warp models using the CMB alone versus the combination of LSS surveys with CMB 
in figures~\ref{fig:kink_CMB}, \ref{fig:step_CMB} and \ref{fig:warp_CMB} respectively.

\end{document}